\documentclass{aa}

\usepackage{graphicx}
\usepackage{txfonts}
\bibpunct{(}{)}{;}{a}{}{,} 

\begin{document} 

   \title{Tracking Advanced Planetary Systems (TAPAS) with HARPS-N} 

   \subtitle{VIII. A wide-orbit planetary companion in the hot-Jupiter system HD~118203
   \thanks{Based on observations obtained with the \textit{Hobby-Eberly} Telescope, which is a joint project of the University of Texas at Austin, the Pennsylvania State University, Stanford University, Ludwig-Maximilians-Universit{\"a}t M{\"u}nchen, and Georg-August-Universit{\"a}t G{\"o}ttingen.}
   \fnmsep
   \thanks{Based on observations made with the Italian Telescopio Nazionale \textit{Galileo} (TNG) operated on the island of La Palma by the Fundaci\'on Galileo Galilei of the INAF (Istituto Nazionale di Astrofisica) at the Spanish Observatorio del Roque de los Muchachos of the Instituto de Astrof\'isica de Canarias.}
   \fnmsep
   \thanks{Based on observations made with the TNG under Director’s Discretionary Time of Spain’s Instituto de Astrof\'isica de Canarias.}}

   \author{G.~Maciejewski\inst{1}
            \and
           A.~Niedzielski\inst{1}
            \and
           K.~Go\'zdziewski\inst{1}
            \and
           A.~Wolszczan\inst{2,3}
            \and
           E.~Villaver\inst{4,5}
            \and 
           M.~Fern\'andez\inst{6}
            \and
           M.~Adam\'ow\inst{7}
            \and
           J.~Sierzputowska\inst{1} 
          }

   \institute{Institute of Astronomy, Faculty of Physics, Astronomy and Informatics,
              Nicolaus Copernicus University, Grudziadzka 5, 87-100 Toru\'n, Poland,
              \email{gmac@umk.pl}
         \and
              Department of Astronomy and Astrophysics, Pennsylvania State University, 525 Davey Laboratory, University Park, PA 16802, USA
         \and
              Center for Exoplanets and Habitable Worlds, Pennsylvania State University, 525 Davey Laboratory, University Park, PA 16802, USA
         \and
             Instituto de Astrof\'isica de Canarias, 38205 La Laguna, Tenerife, Spain
         \and    
             Agencia Espacial Espa\~nola (AEE), 41015 Sevilla, Spain
         \and
             Instituto de Astrof\'isica de Andaluc\'ia (IAA-CSIC), 
             Glorieta de la Astronom\'ia 3, 18008 Granada, Spain
         \and
             Center for AstroPhysical Surveys, National Center for Supercomputing Applications, Urbana, IL 61801, USA
             }
   \authorrunning{G.~Maciejewski et al.}
   \date{Received 12 June 2024 ; accepted 15 July 2024 }
 
 
  \abstract
   {}
   {The star HD 118203, classified as a K0 subgiant, was known to harbour a transiting hot Jupiter planet on a 6.1-day eccentric orbit. Previous studies also revealed a linear trend in the radial velocity (RV) domain, indicative of a companion on a wide orbit. Such a hierarchical orbital architecture could be helpful in studies of the origins of hot Jupiters.}
   {We acquired precise RV measurements over 17 years using the 9.2 m \textit{Hobby-Eberly} Telescope and the 3.6 m Telescopio Nazionale \textit{Galileo}. Combining these observations with space-born photometric time series from the Transiting Exoplanet Survey Satellite, we constructed a two-planetary model for the system. Astrometric observations from \textsc{Hipparcos} and \textit{Gaia} were used to constrain the orbital inclination of the wide-orbit companion and its mass. Numerical simulations were used to investigate the dynamics of the system. The photometric data were searched for additional transit-like flux drops.}
   {We found that the additional companion is an 11-Jupiter mass planet orbiting HD 118203 on a 14-year moderately eccentric orbit, constituting a hierarchical planetary system with the hot Jupiter. Both planets were found to be dynamically decoupled mainly due to the general relativistic apsidal precession of the inner planet, marginalising secular interactions. The orbits of both planets might have a relatively low mutual inclination unless the longitudes of the ascending node differ substantially. This configuration favours the coplanar high-eccentricity migration as a path to the present-day orbital configuration. No other transiting planets with radii down to 2 Earth radii and orbital periods less than 100 days were found in the system.}
   {}


   \maketitle
%

\section{Introduction}\label{Sect:Intro}

Most hot Jupiters are known to be lonely planets in their planetary systems \citep[e.g.][]{2021AJ....162..263H}. Only a select number of them are observed in compact orbital configurations and are accompanied by low-mass planets \citep[e.g.][]{2023MNRAS.524.1113S}. The hot Jupiters are slightly more likely to have cold and massive companions orbiting further away \citep[e.g.][]{2017A&A...602A.107B}. These orbital architectures favour high-eccentricity migration (HEM) as a main formation path for those planets \citep{2017AJ....154..106N}. Under this conception, a Jupiter-like planet is formed beyond the water ice line under the core-accretion mechanism \citep{1980PThPh..64..544M,1996Icar..124...62P}. Then, gravitational interactions with an outer body or bodies, also formed in the system, or with a protoplanetary disk excite an orbital eccentricity with a periapsis not smaller than the planet-star Roche separation. Tidal interactions with the host circularise the orbit, gradually reducing an apoapsis. The planet becomes a hot Jupiter on a tight orbit with a semi-major axis down to twice the Roche separation \citep{2005Icar..175..248F}. This prediction aligns with a distribution of the orbital separations for the vast majority of hot Jupiters observed so far \citep{2017A&A...602A.107B}.

The von Zeipel-Lidov-Kozai\footnote{We follow the naming convention postulated by \citet{2019MEEP....7....1I}} (ZLK) oscillations \citep{1910AN....183..345V,1962P&SS....9..719L,1962AJ.....67..591K} are thought to be one of the possible mechanisms triggering HEM \citep{2003ApJ...589..605W,2011Natur.473..187N}. This scenario assumes a star-planet system is perturbed by a distant third body (a massive planet, brown dwarf, or star). Under some conditions, a cyclic exchange of angular momentum between the inner and outer companion on an inclined orbit occurs, resulting in oscillations in orbital eccentricity and inclination of the inner planet in timescales much longer than its orbital period. When its orbital eccentricity is excited, the inner planet undergoes inward migration driven by the tidal interactions. 

The ZLK mechanism requires mutual inclinations above $\approx 40\degr$ to operate. It can also result in misalignment of the stellar spin and the orbital angular momentum of the hot Jupiter. However, \citet{2017AJ....154..230B} analysed six hot Jupiter systems with exterior companions, finding that the orbital inclinations of those bodies must be aligned within $\approx 20\degr$ with the orbits of the hot Jupiters and the spins of their host stars. The ZLK mechanism is unlikely to shape such well-aligned systems, necessitating the involvement of the coplanar HEM \citep[CHEM,][]{2015ApJ...805...75P}. In that scenario, a hot Jupiter is brought on its tight orbit through the secular gravitational interactions with its more massive companion whose mutual inclination remains below $\approx 20\degr$, the semi-major axis is above $\approx 5$ au, the orbital period is longer than 10 yr, and the eccentricity ranges from 0.2 to 0.5.

In the HD~118203 planetary system, a 2 $M_{\rm Jup}$ planet on a 6.1-day eccentric ($\approx 0.3$) orbit was discovered in the radial velocity (RV) domain within a framework of a metallicity-biased planet-search project by \citet{2006AA...446..717D}. This finding was based on 43 RV measurements acquired between May 2004 and July 2005 with the ELODIE echelle spectrograph \citep{1996A&AS..119..373B} paired with the 1.9 m telescope at the Observatoire de Haute-Provence (OHP). \citet{2020AJ....159..243P} detected the planet's transit signatures in space-borne photometric time series acquired with the Transiting Exoplanet Survey Satellite \citep[TESS,][]{2015JATIS...1a4003R}. Those 3 millimagnitude deep transits were used to derive the planetary radius $R_{\rm b}$ of about $1.1$ $R_{\rm Jup}$ and the orbital inclination $i_{\rm b}$ of about $89\degr$. The host is an 8-magnitude star located in a heliocentric distance of $92.03^{+0.15}_{-0.16}$ pc \citep{2022yCat.1355....0G}. With an effective temperature of 5600 K, a gravity of $\log{g_{\star}} \approx 3.9$ (cgs), mass of $1.2$ $M_{\odot}$, and age of 5 Gyr, it was classified as a subgiant of the K0 spectral type. \citet{2020AJ....159..243P} determined its radius of $2.1$ $R_{\odot}$. Those authors also identified an alternative, less probable solution in which HD~118203 has a mass of $1.5$ $M_{\odot}$ and is 2 Gyr younger. \citet{2024arXiv240117272C} detected a 6.1-day modulation in the TESS photometry. Its periodicity coincides with the orbital period of the hot Jupiter and exhibits variable characteristics. It was attributed to magnetic star-planet interactions, in which stellar activity is coupled to some extent with the planet's orbital motion. 

The analysis of \citet{2006AA...446..717D} revealed a linear RV drift of $49.7 \pm 5.7$ ${\rm m \, s^{-1} \, yr^{-1}}$. It was confirmed by \citet{2020AJ....159..243P} and refined to $50.8 \pm 4.4$ ${\rm m \, s^{-1} \, yr^{-1}}$ using an RV data set extended with 13 unpublished ELODIE measurements acquired between March and June 2006. Speckle imaging eliminated stellar companions fainter by 4--9 mag and located within $1\farcs25$ that could have caused this trend, leaving a low mass star (up to 0.4 $M_{\odot}$ ) or even less massive body in play. This finding, together with the eccentric orbit of HD~118203~b, makes the system an excellent laboratory for studying HEM. The unknown physical properties of the additional companion motivated us to conduct an RV follow-up campaign, leading to the determination of the orbital and physical parameters of the wide-orbit companion.

The rest of this paper is organised as follows. In Sect.~\ref{Sect:Obs}, we present the observational material used in our analysis. A description of this analysis and results are given in Sect.~\ref{Sect:Results}. In Sect.~\ref{Sect:Discussion}, we discuss our findings. Finally, we conclude in Sect.~\ref{Sect:Conclusions}.

\section{Observations}\label{Sect:Obs}

\subsection{Radial velocity}

We acquired 51 RV measurements for the HD~118203 system with the High-Resolution Spectrograph \citep[HRS,][]{1998SPIE.3355..387T} at the 9.2 m \textit{Hobby-Eberly} Telescope \citep[HET,][]{1998SPIE.3352...34R} between January 2006 and June 2013. These observations were performed within the Pennsylvania-Toru\'n Planet Search for exoplanets around evolved stars \citep[PTPS,][]{2007ApJ...669.1354N}. The precise RVs were obtained using the gaseous iodine gas cell technique \citep{1992PASP..104..270M,1996PASP..108..500B} and the IRAF\footnote{IRAF is distributed by the National Optical Astronomy Observatories, which are operated by the Association of Universities for Research in Astronomy, Inc., under cooperative agreement with the National Science Foundation.}-based pipeline, following the procedure described in detail by \citet{2012Nowak} and \citet{2013ApJ...770...53N}.

We also observed the system within the Tracking Advanced Planetary Systems project \citep[TAPAS,][]{2015A&A...573A..36N} between December 2012 and August 2015. A service time request secured an additional observation in March 2023. We used the High Accuracy RV Planet Searcher in the northern hemisphere \citep[HARPS-N,][]{2012SPIE.8446E..1VC} at the 3.6 m Telescopio Nazionale \textit{Galileo} (TNG) for all those observations. In total, we obtained 18 high-precision RV measurements with that facility. The last observation was made almost 8 years after our regular monitoring programme ended. It extended the time coverage of this homogeneous RV data set, which allowed us to achieve the primary goal of our study thanks to the long-term stability of HARPS-N. The RV values were determined with the Data Reduction Software \citep{2002A&A...388..632P, 2007A&A...468.1115L}. In brief, the spectra were wavelength-calibrated using the simultaneous Th-Ar mode of the spectrograph. Then, the RVs and their uncertainties were calculated using the weighted cross-correlation function. The cross-correlation mask for a K5-type star, closest to the spectral type of HD~118203, was used. Tables~\ref{tab.HET} and \ref{tab.HARPSn} show our new RV measurements.

\subsection{TESS photometry}

\begin{figure}
	\centering
	\includegraphics[width=0.8\columnwidth]{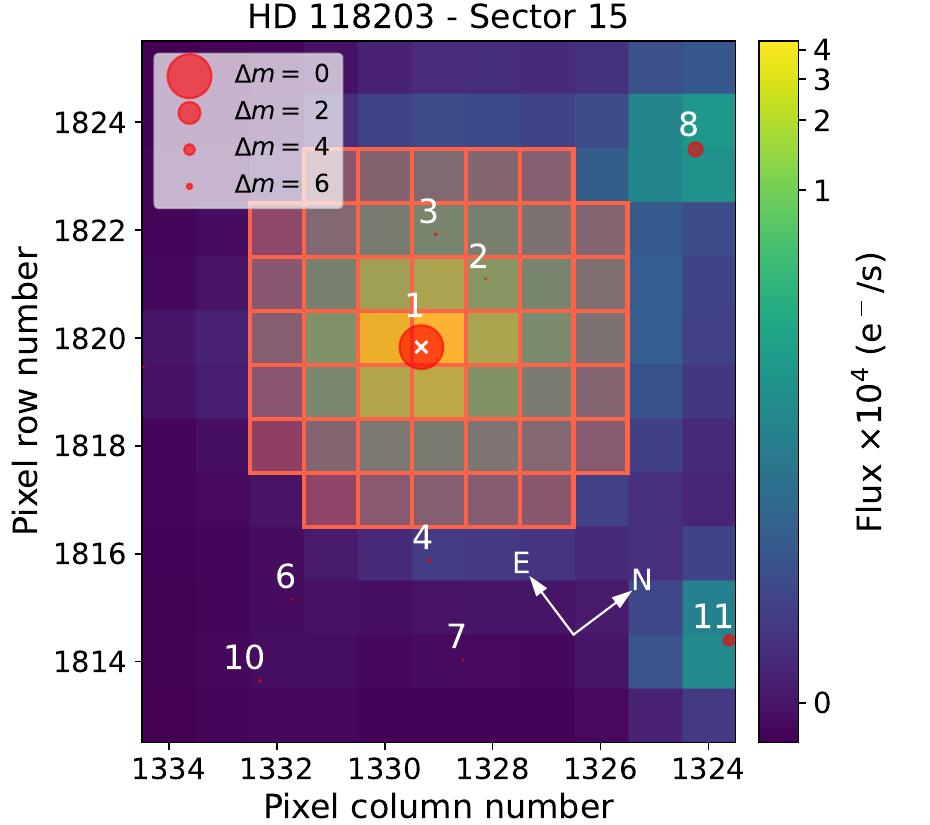}
    \caption{Field of view around HD~118203 observed by TESS in Sector 15. The position of the host star, taken from the \textit{Gaia} DR3 catalogue, is marked near the centre and labelled with 1. Other detected stars are marked with symbols sized according to their catalogue brightness and labelled according to their angular separation from the target. The colour scale reflects fluxes recorded in each pixel. A red overlay shows the custom aperture used for the photometry extraction. The figure was generated with the \texttt{tpfplotter} tool \citep{2020AA...635A.128A}.}
    \label{fig:fov}
\end{figure}

\begin{table*}[h]
\caption{Details on the TESS observations for HD~118203.} 
\label{tab.ObsTESS}      
\centering                  
\begin{tabular}{c c c c c c c}      
\hline\hline                
Sector  & Camera & CCD & from -- to (UT) & $N_{\rm{obs}}$ & PNR (ppth minute$^{-1}$) & $N_{\rm{tr}}$ \\
\hline
15 & 4 & 4 & 2019 Aug 15 to 2019 Sep 11 & 17743 & $0.80$ & 4 \\
16 & 4 & 3 & 2019 Sep 11 to 2019 Oct 07 & 16719 & $0.66$ & 4 \\
22 & 3 & 2 & 2020 Feb 18 to 2020 Mar 18 & 18609 & $0.73$ & 4 \\
49 & 3 & 1 & 2022 Feb 26 to 2022 Mar 26 & 17971 & $0.53$ & 4 \\
76 & 1 & 2 & 2024 Feb 27 to 2024 Mar 25 & 14423 & $0.54$ & 3 \\
\hline                                   
\end{tabular}
\tablefoot{$N_{\rm{obs}}$ is the number of useful data points. PNR is the photometric noise rate  \citep{2011AJ....142...84F} and ppth stands for part per thousand. $N_{\rm{tr}}$ is the number of complete transit light curves used in this study.}
\end{table*}

The HD~118203 system was observed five times during the first six years of TESS operations. We considered only observations acquired in the 2-minute cadence. The details of individual runs are summarised in Table~\ref{tab.ObsTESS}. The photometric time series obtained in the first two sectors in 2019 were used to detect transits for HD~118203~b by \citet{2020AJ....159..243P}. Further observations from early 2020, 2022, and 2024 allowed us to refine the planet's orbital period and other system parameters that can be derived from transit light curve modelling. 

\begin{figure*}
	\includegraphics[width=2\columnwidth]{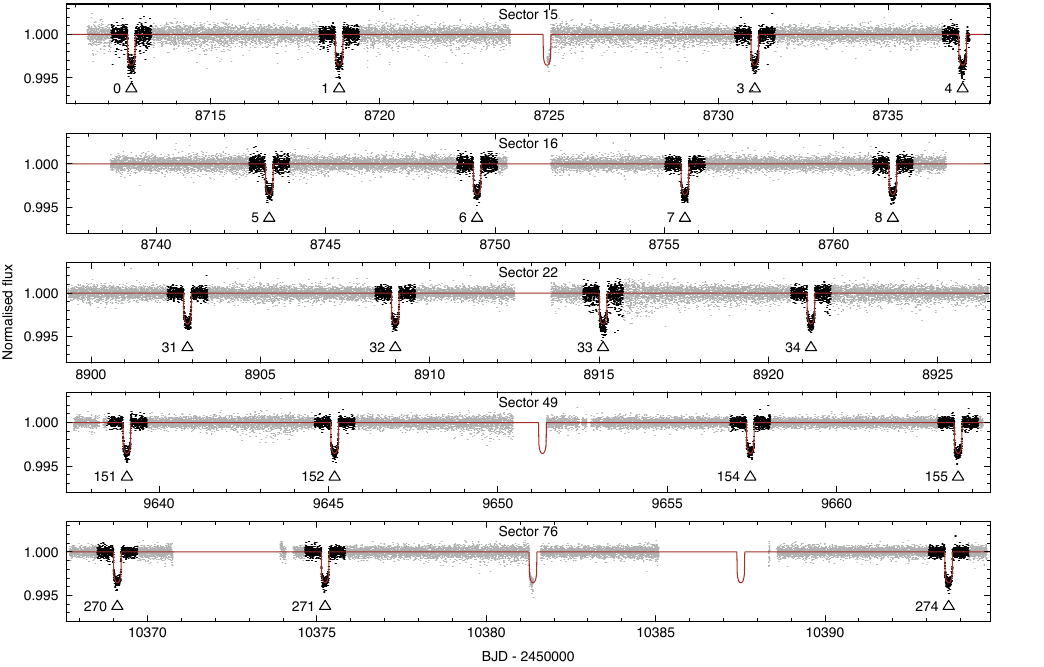}
    \caption{TESS photometric time series extracted for HD~118203. Grey points mark all measurements. The chunks considered in transit modelling for planet~b are plotted with black points. The transits are numbered starting from the reference epoch given by \citet{2020AJ....159..243P}. The red lines sketch the best-fitting transit model derived in Sect.~\ref{SSect:Results.2plSystem}. }
    \label{fig:tesslcs}
\end{figure*}

The fluxes were obtained using standard procedures implemented in the \texttt{Lightkurve v2.0} package \citep{2018ascl.soft12013L}, starting with aperture photometry applied to the target pixel files, available via the exo.MAST portal\footnote{\url{https://exo.mast.stsci.edu}}. The custom aperture mask was iteratively optimised in each sector to produce the lowest photometric noise rate \citep[PNR,][]{2011AJ....142...84F} for a final light curve. Figure~\ref{fig:fov} shows the exemplary TESS pixel map for Sector 15, with HD~118203 near the centre and labelled as 1. The nearby sources found in \textit{Gaia} Data Release 3 \citep[DR3,][]{2023AA...674A...1G} are marked with symbols sized with their brightness and labelled from 2 to 11. In the aperture mask used for the photometry extraction, marked with a red overlay, we found two sources (marked as 2 and 3). Their contributions can be neglected since they are $\ga 9.5$ mag in the $G$ band fainter than the host star. The brightest field stars (marked as 8 and 11), in turn, are located at secure angular separations above $130 \arcsec$ ($6$ pixels), outside the aperture mask in all sectors. Since neglecting their contribution can be justified, we applied no correction for flux contamination. We employed the Savitzky-Golay filter \citep{1964AnaCh..36.1627S} with a window of 12 hours to remove any low-frequency trends from astrophysical or systematic effects. In this step, in-transit data points were masked following a trial transit ephemeris from \citet{2020AJ....159..243P}. The final light curves acquired in the individual sectors are plotted in Fig.~\ref{fig:tesslcs}.

\section{Data analysis and results}\label{Sect:Results}

\subsection{Stellar parameters for HD 118203}\label{SSect:Results.Star}

\begin{table*}[h]
\caption{Physical properties of HD~118203.} 
\label{tab.starParams}      
\centering                  
\begin{tabular}{l c c c c}      
\hline\hline                
Parameter         & da Silva  & Deka-Szymankiewicz & Pepper & Castro-Gonz\'alez \\
    & et al. (2006) & et al. (2018)$^{\rm a)}$ & et al. (2020)$^{\rm b)}$ & et al. (2024)$^{\rm c)}$ \\
\hline
Effective temperature, $T_{\rm{eff}}$ (K)                                                  & $5600\pm150$  & $5788 \pm 51$             & $5683^{+84}_{-85}$        & $5872\pm20$       \\
Surface gravity, $\log g$ (cgs)                                                            & $3.87$        & $3.99 \pm 0.12$           & $3.889^{+0.017}_{-0.018}$ & $4.05 \pm 0.04$   \\
Metallicity, [Fe/H] (dex)                                                                  & $0.10\pm0.05$ & $0.22 \pm 0.02$           & $0.223\pm0.076$           & $0.27 \pm 0.02$   \\
Microturbulence velocity, $v_{\rm{micro}}$ $(\rm{km~s}^{-1})$                              & $...$         & $1.15 \pm 0.18$           & $...$                     & $...$             \\
Projected rotation velocity, $v_{\rm{\star,rot}} \sin{I_{\star}}$ $(\rm{km~s}^{-1})$ & $4.7$         & $3.23 \pm 0.77$           & $...$                     & $...$             \\
Mass, $M_{\star}$ $(M_{\odot})$                                                            & $1.23\pm0.03$ & $1.220 \pm 0.021$         & $1.251^{+0.053}_{-0.059}$ & $1.353 \pm 0.006$ \\
Radius, $R_{\star}$ $(R_{\odot})$                                                          & $...$         & $2.03 \pm 0.16$           & $2.102^{+0.052}_{-0.051}$ & $1.993 \pm 0.065$ \\
Density, $\rho_{\star}$ $(\rho_{\odot})$                                                   & $...$         & $0.155^{+0.004}_{-0.010}$ & $0.135 \pm 0.008$         & $...$             \\
Luminosity, $L_{\star}$ $(L_{\odot})$                                                      & $...$         & $5.01^{+1.01}_{-0.84}$    & $4.15^{+0.35}_{-0.33}$    & $4.42 \pm 0.02$   \\
Age, (Gyr)                                                                                 & $4.6 \pm 0.8$ & $5.25^{+0.36}_{-0.34}$    & $5.32^{+0.96}_{-0.73}$    & $...$             \\
Rotation period, $P_{\rm \star,rot}$ (d)                                             & $...$         & $<32 \pm 8$               & $20 \pm 5$                & $...$             \\
\hline                                   
\end{tabular}
\tablefoot{$^{\rm a)}$ The values reported by \citet{2018AA...615A..31D} were used in this paper. $^{\rm b)}$ There is also a less probable solution with $M_{\star} = 1.481^{+0.045}_{-0.042} \, M_{\odot}$, $R_{\star} = 2.182^{+0.049}_{-0.047} \, R_{\odot}$, $L_{\star} = 4.71^{+0.36}_{-0.33} \, L_{\odot}$, the age of $2.87 \pm 0.31$ Gyr, and other parameters consistent within 1 $\sigma$. See Table~2 in \citet{2020AJ....159..243P} for details. $^{\rm c)}$ The values were taken from \citet{2021AA...656A..53S} and \citet{2023AA...674A...1G}, see Table~2 in \citet{2024arXiv240117272C} for details.}
\end{table*}

We took the physical parameters of HD~118203 from the compilation of \citet{2018AA...615A..31D}. In short words, the effective temperature $T_{\rm{eff}}$, gravitational acceleration at the stellar surface $\log g$, metallicity [Fe/H], and microturbulence velocity $v_{\rm{micro}}$ were determined from HRS spectra using the \texttt{TGVIT} \citep{2005PASJ...57...27T} code. The programme was fed with the equivalent widths of \ion{Fe}{I} and \ion{Fe}{II} spectral lines measured with the \texttt{pyEW} tool \citep{2015A&A...581A..94A}. The projected rotational velocity $v_{\rm{\star,rot}} \sin{I_{\star}}$ was determined also from the HRS spectra using the \texttt{Spectroscopy Made Easy} tool \citep{1996A&AS..118..595V}. The stellar atmospheric parameters and the luminosity based on the \textit{Gaia} parallax \citep{2016A&A...595A...2G} were used to determine the mass $M_{\star}$, the radius $R_{\star}$, and the host star's age. For this purpose, the procedure from \citet{2016A&A...587A.119A} was adopted. It interpolated the theoretical stellar models of \citet{2012MNRAS.427..127B} to determine the parameters with a Bayesian approach.

The values of the stellar parameters used in the further analysis are listed in Table~\ref{tab.starParams}. The literature values are also given for ease of comparison.

\subsection{Two-planet model}\label{SSect:Results.2plSystem}

\begin{figure*}
	\includegraphics[width=2\columnwidth]{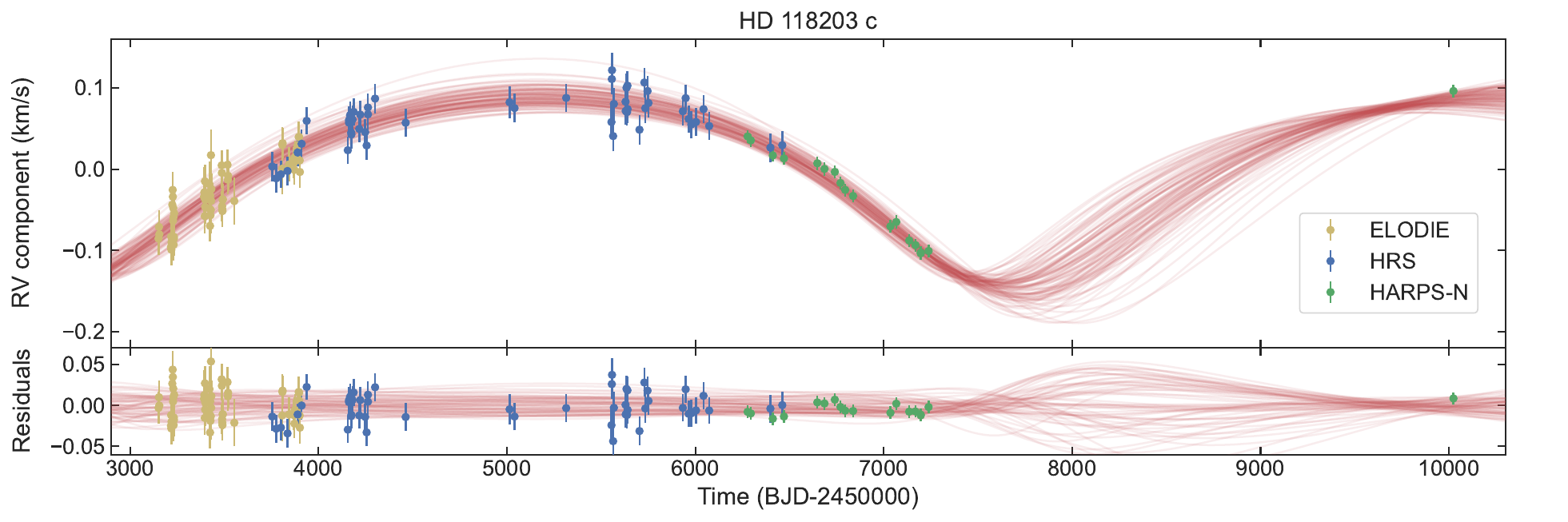}
    \caption{RV reflex motion induced by the wide-orbit companion HD~118203~c. The colours of the data points encode the individual instruments. The uncertainties of the model are illustrated by a bunch of 100 solutions randomly taken from the posterior samples. The residuals between the observed data and the best-fitting model are plotted below.} 
    \label{fig:rv2}
\end{figure*}

Our new RV measurements from 2006 and 2007 confirmed the linear trend reported by \citet{2006AA...446..717D}. As illustrated in Fig.~\ref{fig:rv2}, further observations revealed some curvature and a trend reversal. Our trial models showed that the companion's orbital period and minimal mass remained unconstrained for the upper values, showing a typical correlation between these two parameters. Thus, we revisited the system in 2023 with the precise RV measurement critical for determining the companion's orbital period and other orbital parameters.

We employed the \texttt{ALLESFITTER} tool \citep{allesfitter-code, allesfitter-paper} to construct a model of the three-body system based on the TESS transit light curves for HD~118203~b and the systemic RV data sets. In addition to our HRS and HARPS-N RVs, we used 43 measurements acquired with the ELODIE spectrograph and reported by \citet{2006AA...446..717D}. Following \citet{2020AJ....159..243P}, we extended this data set with 13 additional ELODIE RVs, publicly available from the Data \& Analysis Centre for Exoplanets data repository\footnote{\url{http://dace.unige.ch}}. Those 56 measurements covered three observing seasons from May 2004 to June 2006, partially overlapping with our HRS observations. A summary of the RV data sets used is given in Table~\ref{tab.ObsRV}.

\begin{table*}[h]
\caption{Summary of the RV data sets used in our analysis.} 
\label{tab.ObsRV}      
\centering                  
\begin{tabular}{l l c c c l}      
\hline\hline                
Instrument & Telescope  & from -- to (UT) & $N_{\rm{RV}}$ & $\sigma_{\rm RV}$ $(\rm{km~s}^{-1})$ & Source \\
\hline
ELODIE  & 1.9 m OHP & 2004 May 25 -- 2006 Jun 14 &  56 &  0.0129 & \citet{2006AA...446..717D} \\
HRS     & 9.2 m HET & 2006 Jan 19 -- 2013 Jun 18 &  51 &  0.0083 & This paper \\
HARPS-N & 3.6 m TNG & 2012 Dec 16 -- 2023 Mar 20 &  18 &  0.0017 & This paper \\
\hline                                   
\end{tabular}
\tablefoot{$N_{\rm{RV}}$ is the number of RV measurements. $\sigma_{\rm RV}$ is the median RV precision.}
\end{table*}

We started the modelling procedure by converting all timestamps into barycentric Julian dates and barycentric dynamical time ($\rm{BJD_{TDB}}$). The orbital configurations were coded with the orbital periods $P$, conjunction times $T_0$, RV amplitudes $K$, orbital eccentricities $e$, and arguments of periastron $\omega$. These last two parameters were probed in the space of $\sqrt{e} \cos{\omega}$ and $\sqrt{e} \sin{\omega}$. The RV offset and stellar jitter were allowed to float separately for each RV data set. To model the transits of HD~118203~b, the planet-to-star radii ratio $R_{\rm b}/R_{\star}$, the sum of radii scaled with the semi-major axis $(R_{\rm{b}}+R_{\star})/a_{\rm b}$, and the orbital inclination $i_{\rm b}$ (in the form of $\cos{i_{\rm b}}$) were sampled. The limb darkening (LD) effect, approximated with the quadratic law \citep{1950HarCi.454....1K}, was probed with the $q_1$ and $q_2$ parameters in the transformed $q$-space, as proposed by \citet{2013MNRAS.435.2152K}, and then translated into the physical $\mu$-space represented with the linear $u_1$ and quadratic $u_2$ coefficients. To speed up calculations, the code cut 28-hour-long ($\approx 5$ times the transit duration) chunks around the transit midpoints (see Fig.~\ref{fig:tesslcs}). We considered only transits with complete photometric coverage to provide the most reliable results. Any potential red noise in the time domain was sampled separately for each transit with a Gaussian process based on the Mat\'ern 3/2 kernel.

\begin{figure}
	\includegraphics[width=\columnwidth]{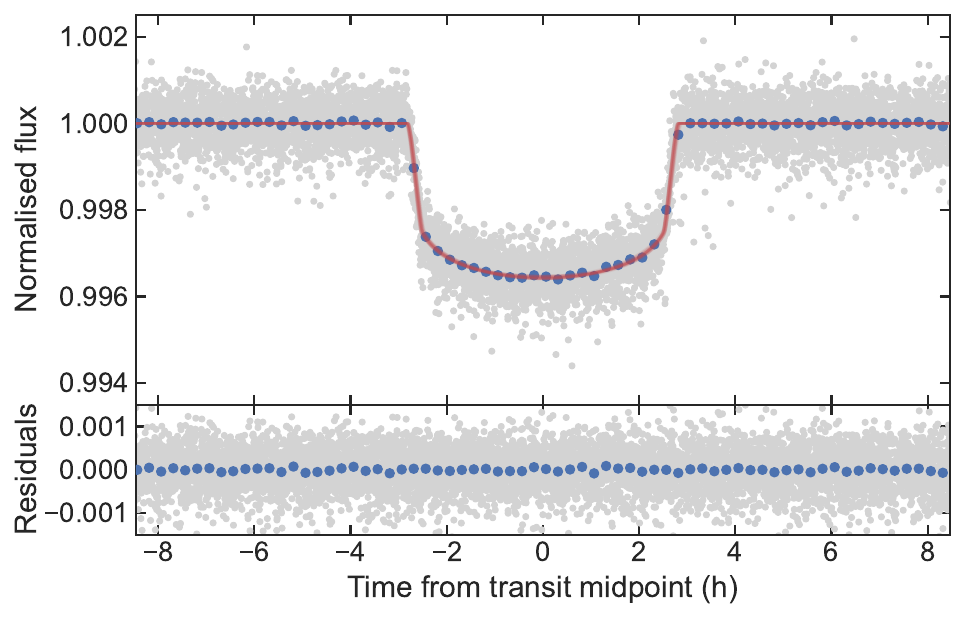}
    \caption{Phase-folded TESS transit light curve for HD~118203~b. Individual measurements are marked with grey points. Blue points show the binned data at 15-minute intervals. The best-fitting model is drawn with a red line. It is blurred with 100 solutions drawn from the posterior samples to reflect the model uncertainties. The residuals are plotted below.}
    \label{fig:transit}
\end{figure}

\begin{figure}
	\includegraphics[width=\columnwidth]{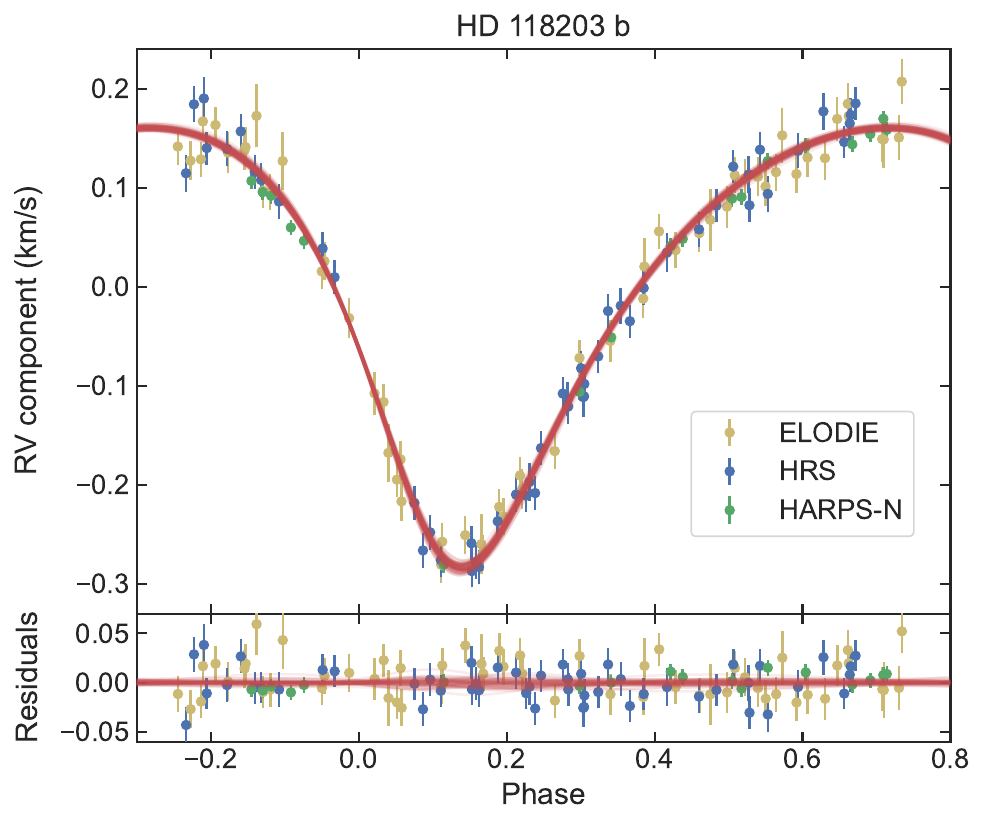}
    \caption{Same as Fig.~\ref{fig:rv2} but for the transiting planet HD~118203~b after phase folding.}
    \label{fig:rv1}
\end{figure}

We used the dynamic nested sampling algorithm with 500 starting living points and the other parameters set as the defaults \citep{allesfitter-paper} to explore the hyperspace of free parameters. The best-fitting model parameters with priors, uncertainties, and calculated physical properties are given in Table~\ref{tab.systemicParams}. The unphased RV signal induced by HD~118203~c, together with the best-fitting model and the residuals, is shown in Fig.~\ref{fig:rv2}. The phase-folded transit light curve and RV component for HD~118203~b are plotted in Figs.~\ref{fig:transit} and \ref{fig:rv1}. 

\subsection{Astrometric constraints on the mass of HD~118203~c}\label{SSect:Results.Astrometry}

\begin{figure}
	\includegraphics[width=\columnwidth]{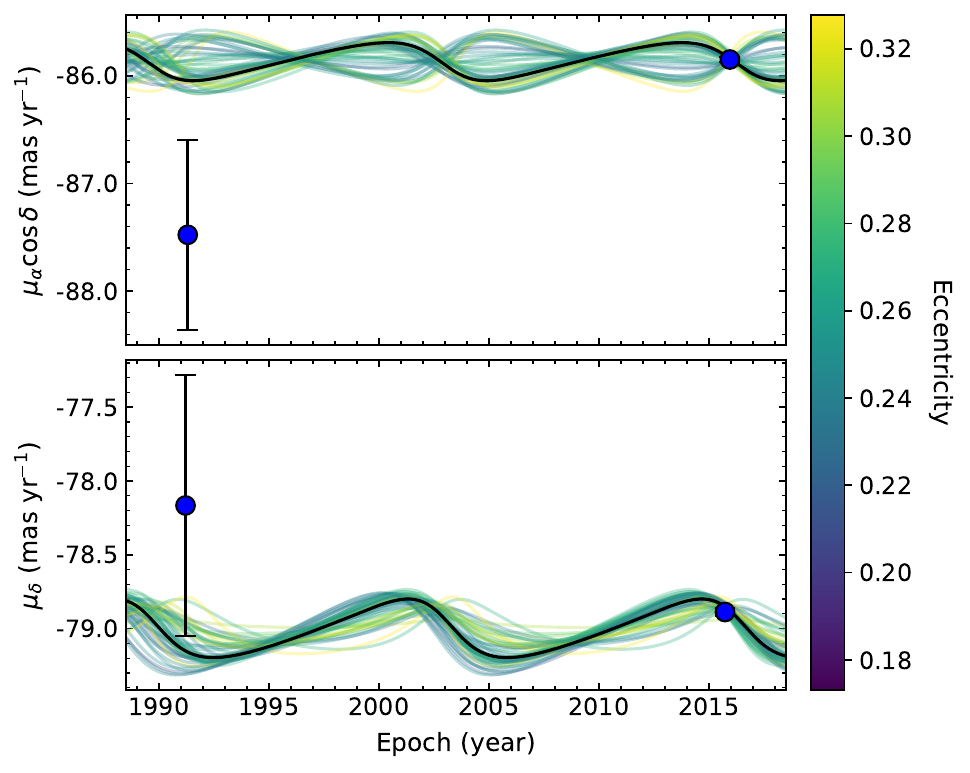}
    \caption{Astrometric accelerations in right ascension ($\mu_{\alpha} \cos \delta$) and declination ($\mu_{\delta}$) for HD~118203. The points near epochs 1991.25 and 2015.8 come from \textsc{Hipparcos} and \textit{Gaia} EDR3, respectively. The best-fitting model is sketched with the black line. The model uncertainties are illustrated with 100 solutions drawn from the posterior distributions. Their colours code the orbital inclination of HD~118203~c, scaled at the right. The figure was plotted with \texttt{ORVARA}.}
    \label{fig:pm}
\end{figure}

We applied \textsc{Hipparcos}-\textit{Gaia} astrometric data to RVs to constrain the value of the orbital inclination of the companion c. We used the \texttt{ORVARA} code \citep{2021AJ....162..186B}, which was fed with the RV component induced by HD~118203~c (Fig.~\ref{fig:rv2}). We simplified a model by subtracting the RV signal of the inner planet and barycentric systemic velocity, considering a two-body system with the central star and the wide-orbit companion. The parameters of the latter, such as $P_{\rm{orb,c}}$, $T_{\rm{0,c}}$, $K_{\rm{c}}$, $e_{\rm{c}}$, and $\omega_{\rm{c}}$, were allowed to vary, starting with priors taken from Sect.~\ref{SSect:Results.2plSystem} (see Table~\ref{tab.systemicParams}). The RV offsets were free for individual instruments. The jitter values reported in Sect.~\ref{SSect:Results.2plSystem} increased the original RV uncertainties. \texttt{ORVARA} combined the RV data with proper motions determined for \textsc{Hipparcos} observations around epoch 1991.25 and \textit{Gaia} around epoch 2015.8. They were extracted from the \textsc{Hipparcos}-\textit{Gaia} Catalogue of Accelerations \citep[][]{2021ApJS..254...42B}, based on the composite \textsc{Hipparcos} catalogue \citep{1997ESASP1200.....E, 2007A&A...474..653V} and \textit{Gaia} Early Third Data Release \citep[EDR3,][]{2021A&A...649A...1G}. The best-fitting model with parameter uncertainties was found with the Markov Chain Monte Carlo method running 200 chains, each $10^5$ steps long. 

The astrometric solution is illustrated in Fig.~\ref{fig:pm}. It constrains the orbital inclination of HD~118203~c, $i_{\rm c}$, to ${95}_{-19}^{+15}$ degrees, demonstrating that the orbital plane of the companion is close to the edge-on configuration. The longitude of its ascending node $\Omega_{\rm c}$ remains unconstrained, uniformly covering a full range of values from 0 to 360 degrees. The remaining orbital parameters were redetermined with a 1-$\sigma$ consistency with the values reported in Sect.~\ref{SSect:Results.2plSystem}, and hence, are skipped in further discussion. The determined value of $i_{\rm c}$ allowed us to calculate the actual mass of HD~118203~c, $M_{\rm c}$, equal to ${11.1}_{-1.0}^{+1.3}$ $M_{\rm{Jup}}$. Thus, the companion could be classified as a massive planet if a deuterium burning criterion \citep[$\approx 13$ $M_{\rm{Jup}}$,][]{2001RvMP...73..719B,2011ApJ...727...57S}, the intersection of the mass distribution of sub-stellar objects \citep[$25$--$30$ $M_{\rm{Jup}}$,][]{2010lyot.confE..11U}, or criteria defined by \citet{2015ApJ...810L..25H}, based on the mass-density-radius distribution, are taken into account. We also note that HD~118203~c is placed at a boundary between low-mass brown dwarfs and super-massive planets if the formation-based boundary definition of $\approx 10$ $M_{\rm{Jup}}$ is considered \citep{2018ApJ...853...37S}.

The proper motions from \textsc{Hipparcos} deviate from the model by 1.6 $\sigma$ in right ascension and 1.1 $\sigma$ in declination. Although this deviation is statistically insignificant, it could hint at an additional body on an orbit much wider than HD~118203~c. Thus, we tested for unresolved binarity using some indicators available in \textit{Gaia} DR3 \citep{2021A&A...649A...2L}. The Renormalised Unit Weight Error (RUWE), which determines whether the single-star model provides a good fit to the astrometric observations, is equal to 0.96, much lower than the cutoff threshold of 1.4 for non-single sources. The amplitude of the variation of the goodness of fit versus position angle of the scan direction (\texttt{ipd$\_$gof$\_$harmonic$\_$amplitude}) and the percentage of observing windows with more than one peak (\texttt{ipd$\_$frac$\_$multi$\_$peak}) also speak in favour of singularity. However, the star shows the excess of astrometric noise $\xi$ of $0.13$ mas (\texttt{astrometric$\_$excess$\_$noise}) with a significance (in terms of the signal-to-noise ratio) of 11 (\texttt{astrometric$\_$excess$\_$noise$\_$sig}). This parameter is the excess uncertainty that must be added in quadrature to obtain a statistically acceptable astrometric solution \citep{2018A&A...616A...1G}. Our model predicts the amplitude of astrometric orbital wobbling $\theta \approx 0.2$ mas, which translates into astrometric noise of $\theta / \sqrt{2} \approx 0.14$ mas, assuming perfect sampling in a simplified circular-orbit approach. Intriguing, this value is close to $\xi$, suggesting that this noise excess is produced by HD~118203~c. Such a scenario opens the opportunity to obtain a complete astrometric solution from raw astrometry, which is expected to be released in \textit{Gaia} DR4.

\subsection{Dynamical analysis}\label{SSect:Results.Dynamical}

\begin{figure*}
 \sidecaption
	\includegraphics[width=12cm]{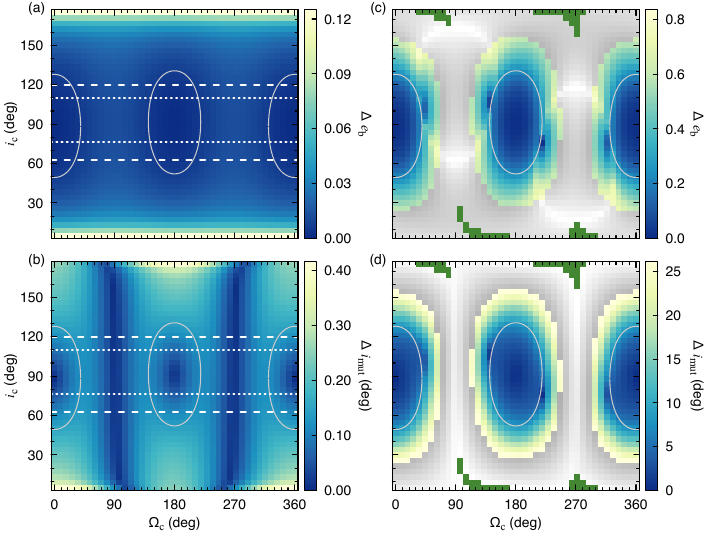}
    \caption{Range of secular variations of $e_{\rm b}$ and $i_{\rm mut}$ from numerical simulations iterated for $i_{\rm c}$ and $\Omega_{\rm c}$. Panels a and b were obtained from the model for which the GR and tidal effects were accounted. The values of those variations, $\Delta e_{\rm b}$ and $\Delta i_{\rm mut}$, are coded with a colour scale. The horizontal dotted and dashed lines show the 1 and 2 $\sigma$ ranges of $i_{\rm c}$ constrained from the astrometric observations. The values of $i_{\rm mut}$ between the Kozai's angles are located outside the ovals marked with light grey continuous lines. Panels c and d show the same parameters but for the Newtonian-only model. The simulations leading to unphysical orbits are plotted in greyscale, and those for which we observed orbit flipping from prograde to retrograde or vice versa are marked in green.}
    \label{fig:maps}
\end{figure*}

The value of $i_{\rm c}$ close to $90\degr$ suggests that the HD~118203 system could be coplanar, consistent with the findings of \citet{2017AJ....154..230B}. However, the difference between the longitudes of ascending nodes for planets b and c, defined as $\Delta \Omega = \Omega_{\rm b} - \Omega_{\rm c}$, remains unconstrained. Hence, the mutual inclination $i_{\rm mut}$ between the orbital planes of both planets,
\begin{equation}
	\cos i_{\rm mut} = \cos {i_{\rm b}} \cos{i_{\rm c}} + \sin {i_{\rm b}} \sin{i_{\rm c}} \cos{\Delta \Omega} \, ,
\end{equation}
remains unknown. We arbitrarily set $\Omega_{\rm b} = 0 \degr$ hereafter. The value of $i_{\rm mut}$ can range between $0\degr$ and $180\degr$, leaving the possibility of substantial non-coplanarity. The ZLK oscillations of pronounced magnitude are expected for $i_{\rm mut}$ between the Kozai angles of about $39\degr$ and $141\degr$ \citep[e.g.][]{2016ARA&A..54..441N}. To investigate whether such scenarios could apply to the HD~118203 system, we run numerical experiments using the \texttt{REBOUND/REBOUNDx 4.0.0} package \citep{2012A&A...537A.128R,2020MNRAS.491.2885T}.

The initial parameters in our simulations were taken from the best-fitting model obtained in Sect.~\ref{SSect:Results.2plSystem} and transformed into Jacobi coordinates \citep{2003ApJ...592.1201L}. We derived the mean anomalies for both planets at the initial epoch of the first transit. We iterated over a full range of $i_{\rm c}$ from $5\degr$ to $175\degr$ and $\Omega_{\rm c}$ from $0\degr$ to $360\degr$ on a $42 \times 42$ grid, adapting the procedure from \citet{2018A&A...619A.115R}. We intentionally did not limit the range of $i_{\rm orb,c}$ to the values constrained in the astrometric analysis (Sect.~\ref{SSect:Results.Astrometry}) in order to draw a complete picture of the system's dynamics. The outer planet's mass was updated according to its simulated inclination, following the $m \sin i$ rule. The approximated timescale of the ZLK oscillations, $\tau_{\rm ZLK}$, was found to be a few $10^5$ yr following Eq.~(1) from \citet{2007ApJ...669.1298F}. Thus, the integration time was set to $2 \times 10^6$ yr to cover several cycles. The time step was set at 0.125~d and 0.1875~d for two dynamical models defined below to ensure adequate sampling during a periapsis passage of planet b and to keep the relative energy error not greater than $10^{-9}$--$10^{-8}$ for the whole integration interval. As the dynamical models, we considered the ``pure'' Newtonian model and its variant with conservative perturbations stemming from General Relativity (GR) and star--planet tides. To follow the orbital evolution of the system for a sufficiently long interval that permits catching essential features of the dynamics in the intermediate time scale, covering a few possible ZLK cycles, we used the symplectic Wisdom-Holman integrator \texttt{WHFast} \citep{1991AJ....102.1528W,2015MNRAS.452..376R}, as well as the fourth-order \texttt{SABA4} \citep{2001CeMDA..80...39L}, which are suited for systems with a mass-dominant central body and relatively small perturbations to the Keplerian orbits. We carefully confirmed the accuracy of numerical integrations in test runs performed for some initial configurations, especially leading to eccentricity excitation as large as 0.99 with the high-order non-symplectic integrator \texttt{IAS15} \citep{2015MNRAS.446.1424R}, which is based on the Gauss--Radau integration scheme of \citet{1985ASSL..115..185E} and offers highly accurate calculations at the expense of higher computing time requirements. We found the results of the simulations to be consistent with each other. 

We used Eq.~(12) from \citet{2009ApJ...698.1778R} to estimate the timescale (period) of apsidal precession due to GR $\tau_{\rm GR}$ for HD~118203~b. We found it shorter than $\tau_{\rm ZLK}$ by one order of magnitude, showing that the GR corrections cannot be neglected. They were incorporated with the \texttt{gr\_potential} implementation. It uses a simplified and numerically efficient perturbing potential, inversely proportional to the square of an orbital separation \citep{1986IAUS..114..105N}, to model the apsidal precession around a mass-dominating central body. In test runs, we compared our results to those obtained with the \texttt{gr} algorithm \citep{1975ApJ...200..221A}, which implements the exact GR model but is much less computationally efficient. Again, we found no noticeable discrepancies between the outcomes of the two dynamical GR models.

We also used Eqs.~(6) and (10) from \citet{2009ApJ...698.1778R} to calculate the precession timescales induced by conservative tidal and rotational deformations implemented in the \texttt{tides\_constant\_time\_lag} extension for \texttt{REBOUNDx}. In the former case, the result was of the order of $\tau_{ZLK}$ and the effect was considered by enabling tidal forces. For simplicity, the stellar spin was assumed to be aligned to the orbital angular momentum. The second-order stellar tidal Love number value, $k_{2,\star}$, was set at 0.0063, interpolated from stellar models provided by \citet{1995A&AS..109..441C}. For the planet, we adopted a Jupiter value of $k_{2,\rm b} = 0.5$ \citep[e.g.][]{2018A&A...613A..32N}. We adopted the value of $P_{\rm \star,rot} = 32 \, {\rm d}$ (Table~\ref{tab.starParams}) as a rough estimate of the stellar rotation period. The rotational timescale, in turn, was found to be an order of magnitude greater and, hence, was skipped in further considerations.

The results of our simulations are summarised in Fig.~\ref{fig:maps}. The range of the secular variations for the orbital eccentricity of HD~118203~b, $\Delta e_{\rm b} = e_{\rm b,max} - e_{\rm b,min}$, where $e_{\rm b,max}$ and $e_{\rm b,min}$ are the maximal and minimal values of $e_{\rm b}$ observed in the simulations, are shown in panel~a. This eccentricity was found to be stable within 0.03 for a wide range of $i_{\rm c}$ and $\Omega_{\rm c}$, including the 2$\sigma$ range of $i_{\rm c}$ derived from astrometry and marked with dashed horizontal lines. Only for $i_{\rm c}$ approaching $0\degr$ or $180\degr$ did we observe $\Delta e_{\rm b}$ as high as 0.15. 

In panel b, we plot the range of the $i_{\rm mut}$ variations, $\Delta i_{\rm mut}$, defined analogously to $\Delta e_{\rm b}$. Again, the observed values were below $0.42\degr$ even for mutual inclinations between the Kozai's angles, which are outside the ovals marked with light grey continuous lines in each panel. These outcomes show that secular interactions between both companions are decoupled. 

Our additional tests showed that this decoupling is caused mainly by relatively fast GR precession and, to a lesser extent, by the tidal effects. To find this out, we performed test simulations for two variants of Newtonian models, one enhanced with the GR effects and another with the tidal effects. The damping of the ZLK oscillations was the order of magnitude weaker in the latter model. \citet{2024A&A...683A.193V} also observed such behaviour for other planetary systems. To illustrate the magnitude of decoupling, we show $\Delta e_{\rm b}$ and $\Delta i_{\rm mut}$ mapped in panels c and d for test simulations performed with Newtonian dynamics only. Here, the ZLK cycles were easily excited for mutual orbital inclinations in the range of Kozai's angles. The values of $e_{\rm b}$ were excited even to unphysical values close to 1, driving the planet onto orbits crossing the Roche limit or colliding with the host star. Such configurations were marked with a grey colour scale on both panels. The variations of $i_{\rm mut}$ increased by almost 2 orders of magnitude compared to the model incorporating the GR and tidal effects. We note that in the case of this Newtonian model, already in a relatively short time coverage of our simulations, it is possible to detect so-called ``flipping'' orbits that are characterised by switching between the prograde and retrograde motion \citep{2011Natur.473..187N,2022AJ....164..232H}. Such orbits are marked with green pixels, close to mutual inclinations of $0\degr$ and $180\degr$ at $\Omega_{\rm c} \simeq 90\degr$ and $270\degr$.

The total precession rate for HD~118203~b was found to be $\dot{\omega}_{\rm b,simulated} \approx 50\arcsec / {\rm yr}$. The predicted amplitude of transit timing variation caused by this effect is $14$ hours throughout the precession cycle of 26000 yr. This precession remains far below the detection threshold in present RV and transit timing data sets.

\subsection{Search for additional transiting planets}\label{SSect:Results.Search4Transits}

\begin{figure}
	\includegraphics[width=\columnwidth]{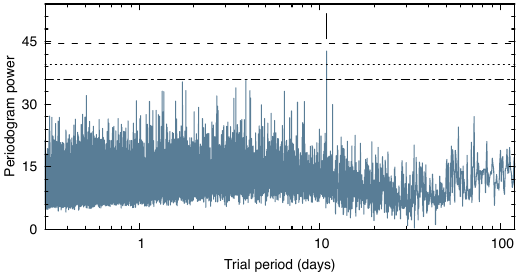}
    \caption{\texttt{AoVtr} periodogram for TESS observations of HD~118203 after cutting out the transits of planet b. A vertical indicator marks the strongest but statistically insignificant peak at 10.9 days. The horizontal lines are the empirical FAP levels of 5\%, 1\%, and 0.1\% (from the bottom up).}
    \label{fig:aovtr}
\end{figure}

We used the TESS photometric time series to search for signatures of additional hypothetical transiting planets in the system. The light curves from individual sectors were merged, and transits of HD~118203~b were masked with 10-minute margins following the updated transit ephemeris. Measurements with a cadence quality flag above 0 were also masked to base the analysis only on reliable data. The Analysis of Variance for planetary Transits \cite[\texttt{AoVtr},][]{2006MNRAS.365..165S}, which is a modification of the Analysis of Variance (AoV) periodogram approach \citep{1989MNRAS.241..153S}, was employed to search for transit-like periodic flux drops. It is based on the periodogram technique optimised to detect box-like periodic signals. A light curve is phase folded and binned for a trial period to test a negative-pulse model with a minimum associated with a transit flux drop. We examined trial periods from 0.3 days, which correspond to an orbital separation equal to $R_{\star}$, to 100 days. A resolution in the frequency domain was set to $5 \times 10^{-5}$ ${\rm d^{-1}}$. Since the \texttt{AoVtr} algorithm is sensitive to the number of bins and the duration of a transit signal, we iterated the bin number from 10 to 200 with a step of 5 to identify a periodogram with the highest peak. As shown in Fig.~\ref{fig:aovtr}, the most substantial peak was found at 10.9 days for 140 bins. To estimate its statistical significance, we applied the bootstrap method. The flux measurements were randomly permuted at the original observing epochs, and periodogram analysis was repeated for $5 \times 10^4$ resampled light curves. The false alarm probabilities (FAPs) were determined from the posterior distribution of the periodogram power of the dominant peaks, which were detected for those resampled light curves. The reference FAP levels of 5\%, 1\%, and 0.1\% are marked with horizontal lines in Fig.~\ref{fig:aovtr}.

The FAP value for the 10.9-day signal was found to be 0.25\%. Although this signal could be formally considered statistically insignificant, the phase-folded light curve was subjected to a thorough visual inspection. An identified flux drop was recognised as being produced by scattered data points rather than an actual transit signal. 

This exercise showed that there are no other transiting planets in the system with periods shorter than 100 d that could be detected using TESS photometry. We performed an injection-recovery test to estimate the upper size limit for hypothetical planets that avoid detection. Artificial transit-like signals, modelled with periodic box-shape flux drops lasting 2 hours, were injected into the original TESS light curve. Our tests showed that most transit-like signals with FAP $<$ 0.1\% could be unquestionably identified by eye in phase-folded light curves \citep{2020AcA....70..181M}. Thus, a transit depth that gave a periodogram peak above a detection threshold of 0.1\% was translated into planetary radius using the value of $R_{\star}$ from Table~\ref{tab.starParams}. This procedure was performed for artificial signals with periods from 0.3 to 100 days. The results of this analysis allowed us to rule out the existence of additional transiting planets greater than 2 $R_{\oplus}$ interior to HD~118203~b and Neptune-like ones in the outer orbits.

\section{Discussion}\label{Sect:Discussion}

Our systemic parameters for HD~118203 agree with the determinations of \citet{2020AJ....159..243P} and \citet{2024arXiv240117272C}, which are well within the 1--2 $\sigma$ range in most cases. One of the exceptions is the transit depth, which was calculated as $(R_{\rm b}/R_{\star})^2$ in the literature. Our value is measured from the model light curve, which is the real value observed in the TESS passband and includes the LD effect. Other exceptions are $a_{\rm{b}}$ and $a_{\rm b}/R_{\star}$, determined by \citet{2024arXiv240117272C}, which deviate by many $\sigma$ from the values reported by us and \citet{2020AJ....159..243P}\footnote{A computational error may have caused these discrepancies, as both parameters were not fitted but derived (A.~Castro-Gonz{\'a}lez, private communication). However, we were unable to confirm this before submission.}. The more extended temporal baseline of the observations resulted in a more precise value of $P_{\rm{orb,b}}$, and the new RV data allowed us to refine other orbital parameters for planet b, such as $e_{\rm{b}}$, $\omega_{\rm{b}}$, and $K_{\rm{b}}$. As \citet{2020AJ....159..243P} predicted, the additional transit light curves marginally improved the transit light curve parameters. While further RV observations would undoubtedly improve our knowledge of the orbital parameters of planet c, our data placed relatively tight constraints on $P_{\rm{orb,c}}$ and $M_{\rm{c}}$. Among cold gas giant planets, that is, with masses or minimal masses between $0.09 \, M_{\rm Jup}$ \citep[$\approx 30~M_{\oplus}$,][]{2015ARA&A..53..409W} and $13 \, M_{\rm Jup}$ \citep{2001RvMP...73..719B,2011ApJ...727...57S}, located on wide orbits, for instance, with orbital periods longer than 10 yr, just about 54\% of them (45 systems identified in the portal exoplanet.eu of The Extrasolar Planets Encyclopaedia) have solidly designated orbital periods with a robust RV coverage. Furthermore, 50\% of them belong to multi-planetary systems. Although the HD~118203 system is not unique in this context, it is still an essential contribution to the field of hot Jupiters with massive companions, as discussed later in this section.

A spurious RV signal may result from asymmetries of spectral lines in stellar spectra. They may arise, for example, from blending of lines, starspots, or stellar pulsations \citep[e.g.][]{2005PASP..117..711G}. In Tables~\ref{tab.HET} and \ref{tab.HARPSn}, we also list the bisector span, which measures the asymmetry of the spectral lines used in radial velocimetry. We found that the long-period RV signal attributed to HD~118203~c does not correlate with the bisectors in either the HRS or HARPS-N data sets. The Pearson correlation coefficients were found to be $r_{\rm HRS} = -0.095$ with a $p$-value of $0.51$ and $r_{\rm HARPS-N} = -0.16$ with a $p$-value of $0.30$, respectively. These results align with the findings of \citet{2024arXiv240117272C}, who used the full width at half maximum and the contrast of the cross-correlation function (CCF) in ELODIE spectra as the stellar activity indicators. No correlations between those indicators and signals that could be attributed to planet b or stellar rotation were observed. A periodic signal found in the CCF contrast was attributed to the sampling window of the ELODIE observations. 

\begin{figure}[h!]
	\includegraphics[width=\columnwidth]{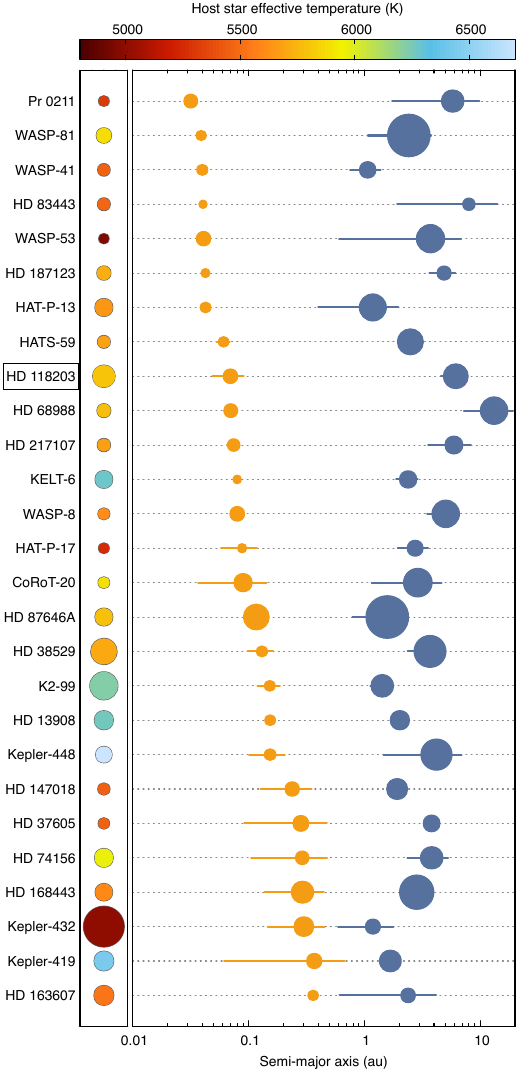}
    \caption{Architectures of hierarchical two-planetary systems with hot or warm gas giants and massive outer companions on wide orbits. The circles in the left panel symbolise the host stars, whose effective temperatures are coded with the colour scale placed at the top of the diagram. Their relative sizes reflect the stars' actual radii. The inner planets and their outer companions are marked in the right panel with mustard and blue-steel circles, respectively. Their sizes are scaled with third-degree roots of the planets' masses or minimal masses. The orbital separations in peri- and apoapsis are shown with horizontal line segments. Note that stars' and planets' sizes are not to the scale. The catalogue data were downloaded from the portal exoplanet.eu of The Extrasolar Planets Encyclopaedia (accessed on 2024 February 2). The HD~118203 system is added separately and marked with a black border. The systems are sorted by the ascending semi-major axis of the inner planets.}
    \label{fig:systems}
\end{figure}

A small value of the ratio of the orbital semi-major axes of both planets orbiting HD~118203, $\alpha = a_{\rm b} / a_{\rm c} = 0.0113 \pm 0.0004$, shows that the system belongs to a class of hierarchical systems for which the semi-major axes of inner planets are much lower than those of the outer companions. In Fig.~\ref{fig:systems}, we plot the known two-planetary systems comprising hot or warm ($P_{\rm orb} < 100 \, {\rm d}$) planets with masses or minimal masses greater than $0.09 \, M_{\rm Jup}$ \citep{2015ARA&A..53..409W} and outer companions on long-period orbits ($P_{\rm orb} > 1 \, {\rm yr}$) and masses or minimum masses above $1 \, M_{\rm Jup}$. We exclude the WASP-47 system in which the hot gas giant is accompanied by two smaller planets \citep{2015ApJ...812L..18B}, constituting a specific compact orbital architecture. We also add the HD~118203 system, in which the outer companion is an $11~M_{\rm Jup}$ planet on a 6 au wide and moderately eccentric orbit. Such orbital architectures could be considered a smoking gun for the hot Jupiter formation path via HEM, mainly if both planets are in eccentric orbits. The system was found to be not far from coplanarity unless there is a substantial difference in the longitudes of the ascending node for the orbits of both companions. Our dynamical simulations demonstrate that HD~118203~b is decoupled from secular interactions of the outer companion, and its orbital evolution is governed by the GR and tidal interactions with the host star. 

HD~118203 is not the only system for which the non-Newtonian effects erased the dynamical memory. We calculated the values of $\tau_{\rm ZLK}$ and $\tau_{\rm GR}$ for the sample of systems collected in Fig.~\ref{fig:systems}. We noticed that among systems with hot gas giants with $a_{\rm b}<0.09$ au (up to HAT-P-17 in Fig.~\ref{fig:systems}), all but HAT-P-13 have $\tau_{\rm GR}$ shorter than $\tau_{\rm ZLK}$. Contrary to this result, we found the reversed relation for all the systems with relatively warm Jupiters on wider orbits (starting with CoRoT-20 in Fig.~\ref{fig:systems}). Thus, in most cases, the hot Jupiters in present-day hierarchical systems seem to be not the best-choice laboratories for studies on the ZLK oscillations. The HAT-P-13 system might be an exception here, with its 0.9 $M_{\rm Jup}$ hot Jupiter and $\geq 14$ $M_{\rm Jup}$ companion on a wide orbit \citep{2009ApJ...707..446B}. The dominance of the GR effects in the hot Jupiter systems, in turn, ensures the long-term dynamic stability of these systems regardless of the mutual inclination of the orbits \citep{2024A&A...683A.193V}.

The dissipation of tides raised in a planet by its host star is the mechanism that circularises its orbit. If the tidal quality factor of the HD~118203~b is $Q_{\rm b} \approx 10^5 - 10^6$, which is attributed to Jupiter \citep{1966Icar....5..375G,1979Natur.279..767Y}, the orbital circularisation time scale $\tau_{\rm b} = - e_{\rm b}/\dot{e}_{\rm b}$ ranges from 1.5 to 14 Gyr (see Eq.~(3) in \citealt{2006ApJ...649.1004A}). The semi-major axis of the final circular orbit is $a_{\rm b,fin} = a_{\rm b} (1-e_{\rm b}^2)$, and the periapsis of the initial orbit is $q_{\rm b,ini} \approx a_{\rm b,fin}/2 \approx 0.031 \, {\rm au}$ \citep{2018ARA&A..56..175D}. If we assume that HD~118203~b was formed ex-situ, beyond the water ice line at the orbital separation of 1--3 au, it did not constitute a hierarchical configuration with the outer companion, and the ZLK mechanism was not a driver of the dynamical evolution. Its initial orbital eccentricity must have been excited to a value as high as 0.97--0.99 to reach the modern short-period orbit. Whether such a configuration could be produced by secular interactions between the two planets under the CHEM scenario requires further studies. On one hand, the orbit of HD~118203~c might have a relatively low mutual inclination, $\lesssim 20\degr$ if $\cos({\Delta \Omega}) \approx 1$ is assumed. Its semi-major axis is wider than 5 au, and the eccentricity is moderately high. \citet{2015ApJ...805...75P} notes that these conditions speak in favour of CHEM. On the other hand, the outer companion is about 5 times more massive than the inner planet, and hence, some fine-tuning of the initial conditions might be required for CHEM to be viable. \citet{2016AJ....152..174A}, in turn, demonstrated that configurations suitable for initiating HEM, in general, could be dynamically unstable for most hot or warm Jupiters with outer companions. Further simulations that are out of the scope of this paper could clarify these issues. 

The TESS data show no sign of additional transiting planets in the system down to $2$ $R_{\oplus}$ for orbits within the orbit of HD~188203~b. For longer periods, the detection threshold increases to about 4 $R_{\oplus}$. The RV residuals against the 2-planet model also show no additional signal. This loneliness of the hot Jupiter, accompanied by the massive companion on the wide orbit, is in line with the arrival of HD~118203~b via the HEM mechanism, which is destructive for close-in and low-massive bodies in a system \citep{2015ApJ...808...14M}.

In the most favourable configuration, the range of transit timing variations for HD~188203~b due to the light travel time effect induced by HD~188203~c would be about 50 s throughout $P_{\rm orb,c}$, which is about 14 years. Such a signal could be detected in precise decade-long transit photometry acquired with presently operating and planned space-borne facilities, such as CHEOPS \citep{2013EPJWC..4703005B} or PLATO \citep{2014ExA....38..249R}. The current timing data set from TESS covers 4.5 years and is too short to detect any changes in the observed value of $P_{\rm orb,b}$. The linear transit ephemeris gives a statistically satisfactory fit with $\chi^2_{\rm reduced}$ of 0.9. Although a trial quadratic ephemeris that might account for a fragment of a periodic signal yields a slightly lower $\chi^2_{\rm reduced}$ of 0.8, it is unequivocally disfavoured by the Bayesian information criterion.

\section{Conclusions}\label{Sect:Conclusions}

The HD~118203 star was found to host a hierarchical planetary system in which a massive planet accompanies a hot Jupiter in an almost 14-year orbit. The newly obtained RVs allowed us to determine the orbital parameters of that companion and, when combined with the astrometric measurements, to determine its mass of about $11~M_{\rm Jup}$. We found its orbital inclination is close to $90\degr$, suggesting that systemic architecture is coplanar as long as there is an unneglectable difference in the longitudes of the orbital node of the planetary orbits. We note that this difference remains, however, unconstrained. Although both planets are decoupled from dynamical secular interactions by the GR and tidal effects, the system may still prove suitable for testing the coplanar high-eccentricity migration. Further Doppler follow-up observations leading to the determination of the degree of alignment or misalignment of the orbital angular momentum and stellar spin through the Rossiter–McLaughlin effect, as well as transit timing and astrometric follow-ups, could shed some light on this issue.

\begin{acknowledgements}
We owe the referee, Dr. Jos\'e A. Caballero, for a prompt response, careful reading, and helpful comments that significantly improved this manuscript. In particular, we thank Him for drawing our attention to the additional binarity indicators available in \textit{Gaia} DR3.
We also thank the HET and IAC resident astronomers and telescope operators for their support. GM acknowledges the financial support from the National Science Centre, Poland, through grant no. 2023/49/B/ST9/00285. 
AN was supported by the Polish National Science Centre grant no. 2015/19/B/ST9/02937. Computations were partly carried out using the computers of Centre of Informatics Tricity Academic Supercomputer \& Network (CI TASK, Poland). 
KG gratefully acknowledges CI TASK for providing support within grant PT01016.
EV acknowledges support from the ``DISCOBOLO'' project funded by the Spanish Ministerio de Ciencia, Innovaci\'on y Universidades under grant PID2021-127289NB-I00. 
MF acknowledges financial support from the Agencia Estatal de
Investigaci\'on (AEI/10.13039/501100011033) of the Ministerio de
Ciencia e Innovaci\'on and the ERDF ``A way of making Europe''
through projects PID2022-137241NB-C43 and the Centre of Excellence
``Severo Ochoa'' award to the Instituto de Astrof\'{\i}sica de
Andaluc\'{\i}a (CEX2021-001131-S).
This paper includes data collected with the TESS mission, obtained from the MAST data archive at the Space Telescope Science Institute (STScI). Funding for the TESS mission is provided by the NASA Explorer Program. STScI is operated by the Association of Universities for Research in Astronomy, Inc., under NASA contract NAS 5-26555. 
The HET is a joint project of the University of Texas at Austin, the Pennsylvania State University, Stanford University, Ludwig-Maximilians-Universit{\"a}t M{\"u}nchen, and Georg-August-Universit{\"a}t G{\"o}ttingen. The HET is named in honor of its principal benefactors, William P. Hobby and Robert E. Eberly. The Center for Exoplanets and Habitable Worlds is supported by the Pennsylvania State University, the Eberly College of Science, and the Pennsylvania Space Grant Consortium.
This research made use of \texttt{Lightkurve}, a Python package for Kepler and TESS data analysis \citep{2018ascl.soft12013L}. 
This work made use of \texttt{tpfplotter} by J. Lillo-Box (publicly available in www.github.com/jlillo/tpfplotter), which also made use of the python packages \texttt{astropy}, \texttt{lightkurve}, \texttt{matplotlib} and \texttt{numpy}. 
This research has made use of data obtained from or tools provided by the portal exoplanet.eu of The Extrasolar Planets Encyclopaedia. This research has made use of the SIMBAD database, operated at CDS, Strasbourg, France. This research has made use of NASA’s Astrophysics Data System.
\end{acknowledgements}

\bibliographystyle{aa} 
\bibliography{hd118} 

\begin{thebibliography}{94}
\expandafter\ifx\csname natexlab\endcsname\relax\def\natexlab#1{#1}\fi

\bibitem[{{Adamczyk} {et~al.}(2016){Adamczyk}, {Deka-Szymankiewicz}, \&
  {Niedzielski}}]{2016A&A...587A.119A}
{Adamczyk}, M., {Deka-Szymankiewicz}, B., \& {Niedzielski}, A. 2016, \aap, 587,
  A119

\bibitem[{{Adam{\'o}w} {et~al.}(2015){Adam{\'o}w}, {Niedzielski}, {Villaver},
  {Wolszczan}, {Kowalik}, {Nowak}, {Adamczyk}, \&
  {Deka-Szymankiewicz}}]{2015A&A...581A..94A}
{Adam{\'o}w}, M., {Niedzielski}, A., {Villaver}, E., {et~al.} 2015, \aap, 581,
  A94

\bibitem[{{Adams} \& {Laughlin}(2006)}]{2006ApJ...649.1004A}
{Adams}, F.~C. \& {Laughlin}, G. 2006, \apj, 649, 1004

\bibitem[{{Aller} {et~al.}(2020){Aller}, {Lillo-Box}, {Jones}, {Miranda}, \&
  {Barcel{\'o} Forteza}}]{2020AA...635A.128A}
{Aller}, A., {Lillo-Box}, J., {Jones}, D., {Miranda}, L.~F., \& {Barcel{\'o}
  Forteza}, S. 2020, \aap, 635, A128

\bibitem[{{Anderson} {et~al.}(1975){Anderson}, {Esposito}, {Martin},
  {Thornton}, \& {Muhleman}}]{1975ApJ...200..221A}
{Anderson}, J.~D., {Esposito}, P.~B., {Martin}, W., {Thornton}, C.~L., \&
  {Muhleman}, D.~O. 1975, \apj, 200, 221

\bibitem[{{Antonini} {et~al.}(2016){Antonini}, {Hamers}, \&
  {Lithwick}}]{2016AJ....152..174A}
{Antonini}, F., {Hamers}, A.~S., \& {Lithwick}, Y. 2016, \aj, 152, 174

\bibitem[{{Bakos} {et~al.}(2009){Bakos}, {Howard}, {Noyes}, {Hartman},
  {Torres}, {Kov{\'a}cs}, {Fischer}, {Latham}, {Johnson}, {Marcy}, {Sasselov},
  {Stefanik}, {Sip{\H{o}}cz}, {Kov{\'a}cs}, {Esquerdo}, {P{\'a}l},
  {L{\'a}z{\'a}r}, {Papp}, \& {S{\'a}ri}}]{2009ApJ...707..446B}
{Bakos}, G.~{\'A}., {Howard}, A.~W., {Noyes}, R.~W., {et~al.} 2009, \apj, 707,
  446

\bibitem[{{Baranne} {et~al.}(1996){Baranne}, {Queloz}, {Mayor}, {Adrianzyk},
  {Knispel}, {Kohler}, {Lacroix}, {Meunier}, {Rimbaud}, \&
  {Vin}}]{1996A&AS..119..373B}
{Baranne}, A., {Queloz}, D., {Mayor}, M., {et~al.} 1996, \aaps, 119, 373

\bibitem[{{Becker} {et~al.}(2017){Becker}, {Vanderburg}, {Adams}, {Khain}, \&
  {Bryan}}]{2017AJ....154..230B}
{Becker}, J.~C., {Vanderburg}, A., {Adams}, F.~C., {Khain}, T., \& {Bryan}, M.
  2017, \aj, 154, 230

\bibitem[{{Becker} {et~al.}(2015){Becker}, {Vanderburg}, {Adams}, {Rappaport},
  \& {Schwengeler}}]{2015ApJ...812L..18B}
{Becker}, J.~C., {Vanderburg}, A., {Adams}, F.~C., {Rappaport}, S.~A., \&
  {Schwengeler}, H.~M. 2015, \apjl, 812, L18

\bibitem[{{Bonomo} {et~al.}(2017){Bonomo}, {Desidera}, {Benatti}, {Borsa},
  {Crespi}, {Damasso}, {Lanza}, {Sozzetti}, {Lodato}, {Marzari}, {Boccato},
  {Claudi}, {Cosentino}, {Covino}, {Gratton}, {Maggio}, {Micela}, {Molinari},
  {Pagano}, {Piotto}, {Poretti}, {Smareglia}, {Affer}, {Biazzo}, {Bignamini},
  {Esposito}, {Giacobbe}, {H{\'e}brard}, {Malavolta}, {Maldonado}, {Mancini},
  {Martinez Fiorenzano}, {Masiero}, {Nascimbeni}, {Pedani}, {Rainer}, \&
  {Scandariato}}]{2017A&A...602A.107B}
{Bonomo}, A.~S., {Desidera}, S., {Benatti}, S., {et~al.} 2017, \aap, 602, A107

\bibitem[{{Brandt}(2021)}]{2021ApJS..254...42B}
{Brandt}, T.~D. 2021, \apjs, 254, 42

\bibitem[{{Brandt} {et~al.}(2021){Brandt}, {Dupuy}, {Li}, {Brandt}, {Zeng},
  {Michalik}, {Bardalez Gagliuffi}, \& {Raposo-Pulido}}]{2021AJ....162..186B}
{Brandt}, T.~D., {Dupuy}, T.~J., {Li}, Y., {et~al.} 2021, \aj, 162, 186

\bibitem[{{Bressan} {et~al.}(2012){Bressan}, {Marigo}, {Girardi}, {Salasnich},
  {Dal Cero}, {Rubele}, \& {Nanni}}]{2012MNRAS.427..127B}
{Bressan}, A., {Marigo}, P., {Girardi}, L., {et~al.} 2012, \mnras, 427, 127

\bibitem[{{Broeg} {et~al.}(2013){Broeg}, {Fortier}, {Ehrenreich}, {Alibert},
  {Baumjohann}, {Benz}, {Deleuil}, {Gillon}, {Ivanov}, {Liseau}, {Meyer},
  {Oloffson}, {Pagano}, {Piotto}, {Pollacco}, {Queloz}, {Ragazzoni}, {Renotte},
  {Steller}, \& {Thomas}}]{2013EPJWC..4703005B}
{Broeg}, C., {Fortier}, A., {Ehrenreich}, D., {et~al.} 2013, in European
  Physical Journal Web of Conferences, Vol.~47, European Physical Journal Web
  of Conferences, 03005

\bibitem[{{Burrows} {et~al.}(2001){Burrows}, {Hubbard}, {Lunine}, \&
  {Liebert}}]{2001RvMP...73..719B}
{Burrows}, A., {Hubbard}, W.~B., {Lunine}, J.~I., \& {Liebert}, J. 2001,
  Reviews of Modern Physics, 73, 719

\bibitem[{{Butler} {et~al.}(1996){Butler}, {Marcy}, {Williams}, {McCarthy},
  {Dosanjh}, \& {Vogt}}]{1996PASP..108..500B}
{Butler}, R.~P., {Marcy}, G.~W., {Williams}, E., {et~al.} 1996, \pasp, 108, 500

\bibitem[{{Castro-Gonz{\'a}lez} {et~al.}(2024){Castro-Gonz{\'a}lez},
  {Lillo-Box}, {Correia}, {Santos}, {Barrado}, {Morales-Calder{\'o}n}, \&
  {Shkolnik}}]{2024arXiv240117272C}
{Castro-Gonz{\'a}lez}, A., {Lillo-Box}, J., {Correia}, A.~C.~M., {et~al.} 2024,
  \aap, 684, A160

\bibitem[{{Claret}(1995)}]{1995A&AS..109..441C}
{Claret}, A. 1995, \aaps, 109, 441

\bibitem[{{Cosentino} {et~al.}(2012){Cosentino}, {Lovis}, {Pepe}, {Collier
  Cameron}, {Latham}, {Molinari}, {Udry}, {Bezawada}, {Black}, {Born},
  {Buchschacher}, {Charbonneau}, {Figueira}, {Fleury}, {Galli}, {Gallie},
  {Gao}, {Ghedina}, {Gonzalez}, {Gonzalez}, {Guerra}, {Henry}, {Horne},
  {Hughes}, {Kelly}, {Lodi}, {Lunney}, {Maire}, {Mayor}, {Micela}, {Ordway},
  {Peacock}, {Phillips}, {Piotto}, {Pollacco}, {Queloz}, {Rice}, {Riverol},
  {Riverol}, {San Juan}, {Sasselov}, {Segransan}, {Sozzetti}, {Sosnowska},
  {Stobie}, {Szentgyorgyi}, {Vick}, \& {Weber}}]{2012SPIE.8446E..1VC}
{Cosentino}, R., {Lovis}, C., {Pepe}, F., {et~al.} 2012, in Society of
  Photo-Optical Instrumentation Engineers (SPIE) Conference Series, Vol. 8446,
  Ground-based and Airborne Instrumentation for Astronomy IV, ed. I.~S.
  {McLean}, S.~K. {Ramsay}, \& H.~{Takami}, 84461V

\bibitem[{{da Silva} {et~al.}(2006){da Silva}, {Udry}, {Bouchy}, {Mayor},
  {Moutou}, {Pont}, {Queloz}, {Santos}, {S{\'e}gransan}, \&
  {Zucker}}]{2006AA...446..717D}
{da Silva}, R., {Udry}, S., {Bouchy}, F., {et~al.} 2006, \aap, 446, 717

\bibitem[{{Dawson} \& {Johnson}(2018)}]{2018ARA&A..56..175D}
{Dawson}, R.~I. \& {Johnson}, J.~A. 2018, \araa, 56, 175

\bibitem[{{Deka-Szymankiewicz} {et~al.}(2018){Deka-Szymankiewicz},
  {Niedzielski}, {Adamczyk}, {Adam{\'o}w}, {Nowak}, \&
  {Wolszczan}}]{2018AA...615A..31D}
{Deka-Szymankiewicz}, B., {Niedzielski}, A., {Adamczyk}, M., {et~al.} 2018,
  \aap, 615, A31

\bibitem[{{ESA}(1997)}]{1997ESASP1200.....E}
{ESA}. 1997, ESA Special Publication, Vol. 1200, {The HIPPARCOS and TYCHO
  catalogues. Astrometric and photometric star catalogues derived from the ESA
  HIPPARCOS Space Astrometry Mission}

\bibitem[{{Everhart}(1985)}]{1985ASSL..115..185E}
{Everhart}, E. 1985, in Astrophysics and Space Science Library, Vol. 115, IAU
  Colloq. 83: Dynamics of Comets: Their Origin and Evolution, ed. A.~{Carusi}
  \& G.~B. {Valsecchi}, 185

\bibitem[{{Faber} {et~al.}(2005){Faber}, {Rasio}, \&
  {Willems}}]{2005Icar..175..248F}
{Faber}, J.~A., {Rasio}, F.~A., \& {Willems}, B. 2005, \icarus, 175, 248

\bibitem[{{Fabrycky} \& {Tremaine}(2007)}]{2007ApJ...669.1298F}
{Fabrycky}, D. \& {Tremaine}, S. 2007, \apj, 669, 1298

\bibitem[{{Fulton} {et~al.}(2011){Fulton}, {Shporer}, {Winn}, {Holman},
  {P{\'a}l}, \& {Gazak}}]{2011AJ....142...84F}
{Fulton}, B.~J., {Shporer}, A., {Winn}, J.~N., {et~al.} 2011, \aj, 142, 84

\bibitem[{{Gaia Collaboration}(2022)}]{2022yCat.1355....0G}
{Gaia Collaboration}. 2022, {VizieR Online Data Catalog: Gaia DR3 Part 1. Main
  source (Gaia Collaboration, 2022)}, VizieR On-line Data Catalog: I/355.
  Originally published in: Astron. Astrophys., in prep. (2022)

\bibitem[{{Gaia Collaboration} {et~al.}(2018){Gaia Collaboration}, {Brown},
  {Vallenari}, {Prusti}, {de Bruijne}, {Babusiaux}, {Bailer-Jones}, {Biermann},
  {Evans}, {Eyer}, {Jansen}, {Jordi}, {Klioner}, {Lammers}, {Lindegren},
  {Luri}, {Mignard}, {Panem}, {Pourbaix}, {Randich}, {Sartoretti}, {Siddiqui},
  {Soubiran}, {van Leeuwen}, {Walton}, {Arenou}, {Bastian}, {Cropper},
  {Drimmel}, {Katz}, {Lattanzi}, {Bakker}, {Cacciari}, {Casta{\~n}eda},
  {Chaoul}, {Cheek}, {De Angeli}, {Fabricius}, {Guerra}, {Holl}, {Masana},
  {Messineo}, {Mowlavi}, {Nienartowicz}, {Panuzzo}, {Portell}, {Riello},
  {Seabroke}, {Tanga}, {Th{\'e}venin}, {Gracia-Abril}, {Comoretto},
  {Garcia-Reinaldos}, {Teyssier}, {Altmann}, {Andrae}, {Audard},
  {Bellas-Velidis}, {Benson}, {Berthier}, {Blomme}, {Burgess}, {Busso},
  {Carry}, {Cellino}, {Clementini}, {Clotet}, {Creevey}, {Davidson}, {De
  Ridder}, {Delchambre}, {Dell'Oro}, {Ducourant},
  {Fern{\'a}ndez-Hern{\'a}ndez}, {Fouesneau}, {Fr{\'e}mat}, {Galluccio},
  {Garc{\'\i}a-Torres}, {Gonz{\'a}lez-N{\'u}{\~n}ez}, {Gonz{\'a}lez-Vidal},
  {Gosset}, {Guy}, {Halbwachs}, {Hambly}, {Harrison}, {Hern{\'a}ndez},
  {Hestroffer}, {Hodgkin}, {Hutton}, {Jasniewicz}, {Jean-Antoine-Piccolo},
  {Jordan}, {Korn}, {Krone-Martins}, {Lanzafame}, {Lebzelter}, {L{\"o}ffler},
  {Manteiga}, {Marrese}, {Mart{\'\i}n-Fleitas}, {Moitinho}, {Mora}, {Muinonen},
  {Osinde}, {Pancino}, {Pauwels}, {Petit}, {Recio-Blanco}, {Richards},
  {Rimoldini}, {Robin}, {Sarro}, {Siopis}, {Smith}, {Sozzetti}, {S{\"u}veges},
  {Torra}, {van Reeven}, {Abbas}, {Abreu Aramburu}, {Accart}, {Aerts},
  {Altavilla}, {{\'A}lvarez}, {Alvarez}, {Alves}, {Anderson}, {Andrei},
  {Anglada Varela}, {Antiche}, {Antoja}, {Arcay}, {Astraatmadja}, {Bach},
  {Baker}, {Balaguer-N{\'u}{\~n}ez}, {Balm}, {Barache}, {Barata}, {Barbato},
  {Barblan}, {Barklem}, {Barrado}, {Barros}, {Barstow}, {Bartholom{\'e}
  Mu{\~n}oz}, {Bassilana}, {Becciani}, {Bellazzini}, {Berihuete}, {Bertone},
  {Bianchi}, {Bienaym{\'e}}, {Blanco-Cuaresma}, {Boch}, {Boeche}, {Bombrun},
  {Borrachero}, {Bossini}, {Bouquillon}, {Bourda}, {Bragaglia}, {Bramante},
  {Breddels}, {Bressan}, {Brouillet}, {Br{\"u}semeister}, {Brugaletta},
  {Bucciarelli}, {Burlacu}, {Busonero}, {Butkevich}, {Buzzi}, {Caffau},
  {Cancelliere}, {Cannizzaro}, {Cantat-Gaudin}, {Carballo}, {Carlucci},
  {Carrasco}, {Casamiquela}, {Castellani}, {Castro-Ginard}, {Charlot},
  {Chemin}, {Chiavassa}, {Cocozza}, {Costigan}, {Cowell}, {Crifo}, {Crosta},
  {Crowley}, {Cuypers}, {Dafonte}, {Damerdji}, {Dapergolas}, {David}, {David},
  {de Laverny}, {De Luise}, {De March}, {de Martino}, {de Souza}, {de Torres},
  {Debosscher}, {del Pozo}, {Delbo}, {Delgado}, {Delgado}, {Di Matteo},
  {Diakite}, {Diener}, {Distefano}, {Dolding}, {Drazinos}, {Dur{\'a}n},
  {Edvardsson}, {Enke}, {Eriksson}, {Esquej}, {Eynard Bontemps}, {Fabre},
  {Fabrizio}, {Faigler}, {Falc{\~a}o}, {Farr{\`a}s Casas}, {Federici},
  {Fedorets}, {Fernique}, {Figueras}, {Filippi}, {Findeisen}, {Fonti},
  {Fraile}, {Fraser}, {Fr{\'e}zouls}, {Gai}, {Galleti}, {Garabato},
  {Garc{\'\i}a-Sedano}, {Garofalo}, {Garralda}, {Gavel}, {Gavras}, {Gerssen},
  {Geyer}, {Giacobbe}, {Gilmore}, {Girona}, {Giuffrida}, {Glass}, {Gomes},
  {Granvik}, {Gueguen}, {Guerrier}, {Guiraud}, {Guti{\'e}rrez-S{\'a}nchez},
  {Haigron}, {Hatzidimitriou}, {Hauser}, {Haywood}, {Heiter}, {Helmi}, {Heu},
  {Hilger}, {Hobbs}, {Hofmann}, {Holland}, {Huckle}, {Hypki}, {Icardi},
  {Jan{\ss}en}, {Jevardat de Fombelle}, {Jonker}, {Juh{\'a}sz}, {Julbe},
  {Karampelas}, {Kewley}, {Klar}, {Kochoska}, {Kohley}, {Kolenberg},
  {Kontizas}, {Kontizas}, {Koposov}, {Kordopatis}, {Kostrzewa-Rutkowska},
  {Koubsky}, {Lambert}, {Lanza}, {Lasne}, {Lavigne}, {Le Fustec}, {Le
  Poncin-Lafitte}, {Lebreton}, {Leccia}, {Leclerc}, {Lecoeur-Taibi},
  {Lenhardt}, {Leroux}, {Liao}, {Licata}, {Lindstr{\o}m}, {Lister}, {Livanou},
  {Lobel}, {L{\'o}pez}, {Managau}, {Mann}, {Mantelet}, {Marchal}, {Marchant},
  {Marconi}, {Marinoni}, {Marschalk{\'o}}, {Marshall}, {Martino}, {Marton},
  {Mary}, {Massari}, {Matijevi{\v{c}}}, {Mazeh}, {McMillan}, {Messina},
  {Michalik}, {Millar}, {Molina}, {Molinaro}, {Moln{\'a}r}, {Montegriffo},
  {Mor}, {Morbidelli}, {Morel}, {Morris}, {Mulone}, {Muraveva}, {Musella},
  {Nelemans}, {Nicastro}, {Noval}, {O'Mullane}, {Ord{\'e}novic},
  {Ord{\'o}{\~n}ez-Blanco}, {Osborne}, {Pagani}, {Pagano}, {Pailler},
  {Palacin}, {Palaversa}, {Panahi}, {Pawlak}, {Piersimoni}, {Pineau}, {Plachy},
  {Plum}, {Poggio}, {Poujoulet}, {Pr{\v{s}}a}, {Pulone}, {Racero}, {Ragaini},
  {Rambaux}, {Ramos-Lerate}, {Regibo}, {Reyl{\'e}}, {Riclet}, {Ripepi}, {Riva},
  {Rivard}, {Rixon}, {Roegiers}, {Roelens}, {Romero-G{\'o}mez}, {Rowell},
  {Royer}, {Ruiz-Dern}, {Sadowski}, {Sagrist{\`a} Sell{\'e}s}, {Sahlmann},
  {Salgado}, {Salguero}, {Sanna}, {Santana-Ros}, {Sarasso}, {Savietto},
  {Schultheis}, {Sciacca}, {Segol}, {Segovia}, {S{\'e}gransan}, {Shih},
  {Siltala}, {Silva}, {Smart}, {Smith}, {Solano}, {Solitro}, {Sordo}, {Soria
  Nieto}, {Souchay}, {Spagna}, {Spoto}, {Stampa}, {Steele},
  {Steidelm{\"u}ller}, {Stephenson}, {Stoev}, {Suess}, {Surdej}, {Szabados},
  {Szegedi-Elek}, {Tapiador}, {Taris}, {Tauran}, {Taylor}, {Teixeira},
  {Terrett}, {Teyssandier}, {Thuillot}, {Titarenko}, {Torra Clotet}, {Turon},
  {Ulla}, {Utrilla}, {Uzzi}, {Vaillant}, {Valentini}, {Valette}, {van Elteren},
  {Van Hemelryck}, {van Leeuwen}, {Vaschetto}, {Vecchiato}, {Veljanoski},
  {Viala}, {Vicente}, {Vogt}, {von Essen}, {Voss}, {Votruba}, {Voutsinas},
  {Walmsley}, {Weiler}, {Wertz}, {Wevers}, {Wyrzykowski}, {Yoldas},
  {{\v{Z}}erjal}, {Ziaeepour}, {Zorec}, {Zschocke}, {Zucker}, {Zurbach}, \&
  {Zwitter}}]{2018A&A...616A...1G}
{Gaia Collaboration}, {Brown}, A.~G.~A., {Vallenari}, A., {et~al.} 2018, \aap,
  616, A1

\bibitem[{{Gaia Collaboration} {et~al.}(2021){Gaia Collaboration}, {Brown},
  {Vallenari}, {Prusti}, {de Bruijne}, {Babusiaux}, {Biermann}, {Creevey},
  {Evans}, {Eyer}, {Hutton}, {Jansen}, {Jordi}, {Klioner}, {Lammers},
  {Lindegren}, {Luri}, {Mignard}, {Panem}, {Pourbaix}, {Randich}, {Sartoretti},
  {Soubiran}, {Walton}, {Arenou}, {Bailer-Jones}, {Bastian}, {Cropper},
  {Drimmel}, {Katz}, {Lattanzi}, {van Leeuwen}, {Bakker}, {Cacciari},
  {Casta{\~n}eda}, {De Angeli}, {Ducourant}, {Fabricius}, {Fouesneau},
  {Fr{\'e}mat}, {Guerra}, {Guerrier}, {Guiraud}, {Jean-Antoine Piccolo},
  {Masana}, {Messineo}, {Mowlavi}, {Nicolas}, {Nienartowicz}, {Pailler},
  {Panuzzo}, {Riclet}, {Roux}, {Seabroke}, {Sordo}, {Tanga}, {Th{\'e}venin},
  {Gracia-Abril}, {Portell}, {Teyssier}, {Altmann}, {Andrae}, {Bellas-Velidis},
  {Benson}, {Berthier}, {Blomme}, {Brugaletta}, {Burgess}, {Busso}, {Carry},
  {Cellino}, {Cheek}, {Clementini}, {Damerdji}, {Davidson}, {Delchambre},
  {Dell'Oro}, {Fern{\'a}ndez-Hern{\'a}ndez}, {Galluccio}, {Garc{\'\i}a-Lario},
  {Garcia-Reinaldos}, {Gonz{\'a}lez-N{\'u}{\~n}ez}, {Gosset}, {Haigron},
  {Halbwachs}, {Hambly}, {Harrison}, {Hatzidimitriou}, {Heiter},
  {Hern{\'a}ndez}, {Hestroffer}, {Hodgkin}, {Holl}, {Jan{\ss}en}, {Jevardat de
  Fombelle}, {Jordan}, {Krone-Martins}, {Lanzafame}, {L{\"o}ffler}, {Lorca},
  {Manteiga}, {Marchal}, {Marrese}, {Moitinho}, {Mora}, {Muinonen}, {Osborne},
  {Pancino}, {Pauwels}, {Petit}, {Recio-Blanco}, {Richards}, {Riello},
  {Rimoldini}, {Robin}, {Roegiers}, {Rybizki}, {Sarro}, {Siopis}, {Smith},
  {Sozzetti}, {Ulla}, {Utrilla}, {van Leeuwen}, {van Reeven}, {Abbas}, {Abreu
  Aramburu}, {Accart}, {Aerts}, {Aguado}, {Ajaj}, {Altavilla}, {{\'A}lvarez},
  {{\'A}lvarez Cid-Fuentes}, {Alves}, {Anderson}, {Anglada Varela}, {Antoja},
  {Audard}, {Baines}, {Baker}, {Balaguer-N{\'u}{\~n}ez}, {Balbinot}, {Balog},
  {Barache}, {Barbato}, {Barros}, {Barstow}, {Bartolom{\'e}}, {Bassilana},
  {Bauchet}, {Baudesson-Stella}, {Becciani}, {Bellazzini}, {Bernet}, {Bertone},
  {Bianchi}, {Blanco-Cuaresma}, {Boch}, {Bombrun}, {Bossini}, {Bouquillon},
  {Bragaglia}, {Bramante}, {Breedt}, {Bressan}, {Brouillet}, {Bucciarelli},
  {Burlacu}, {Busonero}, {Butkevich}, {Buzzi}, {Caffau}, {Cancelliere},
  {C{\'a}novas}, {Cantat-Gaudin}, {Carballo}, {Carlucci}, {Carnerero},
  {Carrasco}, {Casamiquela}, {Castellani}, {Castro-Ginard}, {Castro Sampol},
  {Chaoul}, {Charlot}, {Chemin}, {Chiavassa}, {Cioni}, {Comoretto}, {Cooper},
  {Cornez}, {Cowell}, {Crifo}, {Crosta}, {Crowley}, {Dafonte}, {Dapergolas},
  {David}, {David}, {de Laverny}, {De Luise}, {De March}, {De Ridder}, {de
  Souza}, {de Teodoro}, {de Torres}, {del Peloso}, {del Pozo}, {Delbo},
  {Delgado}, {Delgado}, {Delisle}, {Di Matteo}, {Diakite}, {Diener},
  {Distefano}, {Dolding}, {Eappachen}, {Edvardsson}, {Enke}, {Esquej}, {Fabre},
  {Fabrizio}, {Faigler}, {Fedorets}, {Fernique}, {Fienga}, {Figueras},
  {Fouron}, {Fragkoudi}, {Fraile}, {Franke}, {Gai}, {Garabato},
  {Garcia-Gutierrez}, {Garc{\'\i}a-Torres}, {Garofalo}, {Gavras}, {Gerlach},
  {Geyer}, {Giacobbe}, {Gilmore}, {Girona}, {Giuffrida}, {Gomel}, {Gomez},
  {Gonzalez-Santamaria}, {Gonz{\'a}lez-Vidal}, {Granvik},
  {Guti{\'e}rrez-S{\'a}nchez}, {Guy}, {Hauser}, {Haywood}, {Helmi}, {Hidalgo},
  {Hilger}, {H{\l}adczuk}, {Hobbs}, {Holland}, {Huckle}, {Jasniewicz},
  {Jonker}, {Juaristi Campillo}, {Julbe}, {Karbevska}, {Kervella}, {Khanna},
  {Kochoska}, {Kontizas}, {Kordopatis}, {Korn}, {Kostrzewa-Rutkowska},
  {Kruszy{\'n}ska}, {Lambert}, {Lanza}, {Lasne}, {Le Campion}, {Le Fustec},
  {Lebreton}, {Lebzelter}, {Leccia}, {Leclerc}, {Lecoeur-Taibi}, {Liao},
  {Licata}, {Lindstr{\o}m}, {Lister}, {Livanou}, {Lobel}, {Madrero Pardo},
  {Managau}, {Mann}, {Marchant}, {Marconi}, {Marcos Santos}, {Marinoni},
  {Marocco}, {Marshall}, {Martin Polo}, {Mart{\'\i}n-Fleitas}, {Masip},
  {Massari}, {Mastrobuono-Battisti}, {Mazeh}, {McMillan}, {Messina},
  {Michalik}, {Millar}, {Mints}, {Molina}, {Molinaro}, {Moln{\'a}r},
  {Montegriffo}, {Mor}, {Morbidelli}, {Morel}, {Morris}, {Mulone}, {Munoz},
  {Muraveva}, {Murphy}, {Musella}, {Noval}, {Ord{\'e}novic}, {Orr{\`u}},
  {Osinde}, {Pagani}, {Pagano}, {Palaversa}, {Palicio}, {Panahi}, {Pawlak},
  {Pe{\~n}alosa Esteller}, {Penttil{\"a}}, {Piersimoni}, {Pineau}, {Plachy},
  {Plum}, {Poggio}, {Poretti}, {Poujoulet}, {Pr{\v{s}}a}, {Pulone}, {Racero},
  {Ragaini}, {Rainer}, {Raiteri}, {Rambaux}, {Ramos}, {Ramos-Lerate}, {Re
  Fiorentin}, {Regibo}, {Reyl{\'e}}, {Ripepi}, {Riva}, {Rixon}, {Robichon},
  {Robin}, {Roelens}, {Rohrbasser}, {Romero-G{\'o}mez}, {Rowell}, {Royer},
  {Rybicki}, {Sadowski}, {Sagrist{\`a} Sell{\'e}s}, {Sahlmann}, {Salgado},
  {Salguero}, {Samaras}, {Sanchez Gimenez}, {Sanna}, {Santove{\~n}a},
  {Sarasso}, {Schultheis}, {Sciacca}, {Segol}, {Segovia}, {S{\'e}gransan},
  {Semeux}, {Shahaf}, {Siddiqui}, {Siebert}, {Siltala}, {Slezak}, {Smart},
  {Solano}, {Solitro}, {Souami}, {Souchay}, {Spagna}, {Spoto}, {Steele},
  {Steidelm{\"u}ller}, {Stephenson}, {S{\"u}veges}, {Szabados}, {Szegedi-Elek},
  {Taris}, {Tauran}, {Taylor}, {Teixeira}, {Thuillot}, {Tonello}, {Torra},
  {Torra}, {Turon}, {Unger}, {Vaillant}, {van Dillen}, {Vanel}, {Vecchiato},
  {Viala}, {Vicente}, {Voutsinas}, {Weiler}, {Wevers}, {Wyrzykowski}, {Yoldas},
  {Yvard}, {Zhao}, {Zorec}, {Zucker}, {Zurbach}, \&
  {Zwitter}}]{2021A&A...649A...1G}
{Gaia Collaboration}, {Brown}, A.~G.~A., {Vallenari}, A., {et~al.} 2021, \aap,
  649, A1

\bibitem[{{Gaia Collaboration} {et~al.}(2016){Gaia Collaboration}, {Brown},
  {Vallenari}, {Prusti}, {de Bruijne}, {Mignard}, {Drimmel}, {Babusiaux},
  {Bailer-Jones}, {Bastian}, {Biermann}, {Evans}, {Eyer}, {Jansen}, {Jordi},
  {Katz}, {Klioner}, {Lammers}, {Lindegren}, {Luri}, {O'Mullane}, {Panem},
  {Pourbaix}, {Randich}, {Sartoretti}, {Siddiqui}, {Soubiran}, {Valette}, {van
  Leeuwen}, {Walton}, {Aerts}, {Arenou}, {Cropper}, {H{\o}g}, {Lattanzi},
  {Grebel}, {Holland}, {Huc}, {Passot}, {Perryman}, {Bramante}, {Cacciari},
  {Casta{\~n}eda}, {Chaoul}, {Cheek}, {De Angeli}, {Fabricius}, {Guerra},
  {Hern{\'a}ndez}, {Jean-Antoine-Piccolo}, {Masana}, {Messineo}, {Mowlavi},
  {Nienartowicz}, {Ord{\'o}{\~n}ez-Blanco}, {Panuzzo}, {Portell}, {Richards},
  {Riello}, {Seabroke}, {Tanga}, {Th{\'e}venin}, {Torra}, {Els},
  {Gracia-Abril}, {Comoretto}, {Garcia-Reinaldos}, {Lock}, {Mercier},
  {Altmann}, {Andrae}, {Astraatmadja}, {Bellas-Velidis}, {Benson}, {Berthier},
  {Blomme}, {Busso}, {Carry}, {Cellino}, {Clementini}, {Cowell}, {Creevey},
  {Cuypers}, {Davidson}, {De Ridder}, {de Torres}, {Delchambre}, {Dell'Oro},
  {Ducourant}, {Fr{\'e}mat}, {Garc{\'\i}a-Torres}, {Gosset}, {Halbwachs},
  {Hambly}, {Harrison}, {Hauser}, {Hestroffer}, {Hodgkin}, {Huckle}, {Hutton},
  {Jasniewicz}, {Jordan}, {Kontizas}, {Korn}, {Lanzafame}, {Manteiga},
  {Moitinho}, {Muinonen}, {Osinde}, {Pancino}, {Pauwels}, {Petit},
  {Recio-Blanco}, {Robin}, {Sarro}, {Siopis}, {Smith}, {Smith}, {Sozzetti},
  {Thuillot}, {van Reeven}, {Viala}, {Abbas}, {Abreu Aramburu}, {Accart},
  {Aguado}, {Allan}, {Allasia}, {Altavilla}, {{\'A}lvarez}, {Alves},
  {Anderson}, {Andrei}, {Anglada Varela}, {Antiche}, {Antoja}, {Ant{\'o}n},
  {Arcay}, {Bach}, {Baker}, {Balaguer-N{\'u}{\~n}ez}, {Barache}, {Barata},
  {Barbier}, {Barblan}, {Barrado y Navascu{\'e}s}, {Barros}, {Barstow},
  {Becciani}, {Bellazzini}, {Bello Garc{\'\i}a}, {Belokurov}, {Bendjoya},
  {Berihuete}, {Bianchi}, {Bienaym{\'e}}, {Billebaud}, {Blagorodnova},
  {Blanco-Cuaresma}, {Boch}, {Bombrun}, {Borrachero}, {Bouquillon}, {Bourda},
  {Bouy}, {Bragaglia}, {Breddels}, {Brouillet}, {Br{\"u}semeister},
  {Bucciarelli}, {Burgess}, {Burgon}, {Burlacu}, {Busonero}, {Buzzi}, {Caffau},
  {Cambras}, {Campbell}, {Cancelliere}, {Cantat-Gaudin}, {Carlucci},
  {Carrasco}, {Castellani}, {Charlot}, {Charnas}, {Chiavassa}, {Clotet},
  {Cocozza}, {Collins}, {Costigan}, {Crifo}, {Cross}, {Crosta}, {Crowley},
  {Dafonte}, {Damerdji}, {Dapergolas}, {David}, {David}, {De Cat}, {de Felice},
  {de Laverny}, {De Luise}, {De March}, {de Martino}, {de Souza}, {Debosscher},
  {del Pozo}, {Delbo}, {Delgado}, {Delgado}, {Di Matteo}, {Diakite},
  {Distefano}, {Dolding}, {Dos Anjos}, {Drazinos}, {Duran}, {Dzigan},
  {Edvardsson}, {Enke}, {Evans}, {Eynard Bontemps}, {Fabre}, {Fabrizio},
  {Faigler}, {Falc{\~a}o}, {Farr{\`a}s Casas}, {Federici}, {Fedorets},
  {Fern{\'a}ndez-Hern{\'a}ndez}, {Fernique}, {Fienga}, {Figueras}, {Filippi},
  {Findeisen}, {Fonti}, {Fouesneau}, {Fraile}, {Fraser}, {Fuchs}, {Gai},
  {Galleti}, {Galluccio}, {Garabato}, {Garc{\'\i}a-Sedano}, {Garofalo},
  {Garralda}, {Gavras}, {Gerssen}, {Geyer}, {Gilmore}, {Girona}, {Giuffrida},
  {Gomes}, {Gonz{\'a}lez-Marcos}, {Gonz{\'a}lez-N{\'u}{\~n}ez},
  {Gonz{\'a}lez-Vidal}, {Granvik}, {Guerrier}, {Guillout}, {Guiraud},
  {G{\'u}rpide}, {Guti{\'e}rrez-S{\'a}nchez}, {Guy}, {Haigron},
  {Hatzidimitriou}, {Haywood}, {Heiter}, {Helmi}, {Hobbs}, {Hofmann}, {Holl},
  {Holland}, {Hunt}, {Hypki}, {Icardi}, {Irwin}, {Jevardat de Fombelle},
  {Jofr{\'e}}, {Jonker}, {Jorissen}, {Julbe}, {Karampelas}, {Kochoska},
  {Kohley}, {Kolenberg}, {Kontizas}, {Koposov}, {Kordopatis}, {Koubsky},
  {Krone-Martins}, {Kudryashova}, {Kull}, {Bachchan}, {Lacoste-Seris}, {Lanza},
  {Lavigne}, {Le Poncin-Lafitte}, {Lebreton}, {Lebzelter}, {Leccia}, {Leclerc},
  {Lecoeur-Taibi}, {Lemaitre}, {Lenhardt}, {Leroux}, {Liao}, {Licata},
  {Lindstr{\o}m}, {Lister}, {Livanou}, {Lobel}, {L{\"o}ffler}, {L{\'o}pez},
  {Lorenz}, {MacDonald}, {Magalh{\~a}es Fernandes}, {Managau}, {Mann},
  {Mantelet}, {Marchal}, {Marchant}, {Marconi}, {Marinoni}, {Marrese},
  {Marschalk{\'o}}, {Marshall}, {Mart{\'\i}n-Fleitas}, {Martino}, {Mary},
  {Matijevi{\v{c}}}, {Mazeh}, {McMillan}, {Messina}, {Michalik}, {Millar},
  {Miranda}, {Molina}, {Molinaro}, {Molinaro}, {Moln{\'a}r}, {Moniez},
  {Montegriffo}, {Mor}, {Mora}, {Morbidelli}, {Morel}, {Morgenthaler},
  {Morris}, {Mulone}, {Muraveva}, {Musella}, {Narbonne}, {Nelemans},
  {Nicastro}, {Noval}, {Ord{\'e}novic}, {Ordieres-Mer{\'e}}, {Osborne},
  {Pagani}, {Pagano}, {Pailler}, {Palacin}, {Palaversa}, {Parsons}, {Pecoraro},
  {Pedrosa}, {Pentik{\"a}inen}, {Pichon}, {Piersimoni}, {Pineau}, {Plachy},
  {Plum}, {Poujoulet}, {Pr{\v{s}}a}, {Pulone}, {Ragaini}, {Rago}, {Rambaux},
  {Ramos-Lerate}, {Ranalli}, {Rauw}, {Read}, {Regibo}, {Reyl{\'e}}, {Ribeiro},
  {Rimoldini}, {Ripepi}, {Riva}, {Rixon}, {Roelens}, {Romero-G{\'o}mez},
  {Rowell}, {Royer}, {Ruiz-Dern}, {Sadowski}, {Sagrist{\`a} Sell{\'e}s},
  {Sahlmann}, {Salgado}, {Salguero}, {Sarasso}, {Savietto}, {Schultheis},
  {Sciacca}, {Segol}, {Segovia}, {Segransan}, {Shih}, {Smareglia}, {Smart},
  {Solano}, {Solitro}, {Sordo}, {Soria Nieto}, {Souchay}, {Spagna}, {Spoto},
  {Stampa}, {Steele}, {Steidelm{\"u}ller}, {Stephenson}, {Stoev}, {Suess},
  {S{\"u}veges}, {Surdej}, {Szabados}, {Szegedi-Elek}, {Tapiador}, {Taris},
  {Tauran}, {Taylor}, {Teixeira}, {Terrett}, {Tingley}, {Trager}, {Turon},
  {Ulla}, {Utrilla}, {Valentini}, {van Elteren}, {Van Hemelryck}, {van
  Leeuwen}, {Varadi}, {Vecchiato}, {Veljanoski}, {Via}, {Vicente}, {Vogt},
  {Voss}, {Votruba}, {Voutsinas}, {Walmsley}, {Weiler}, {Weingrill}, {Wevers},
  {Wyrzykowski}, {Yoldas}, {{\v{Z}}erjal}, {Zucker}, {Zurbach}, {Zwitter},
  {Alecu}, {Allen}, {Allende Prieto}, {Amorim}, {Anglada-Escud{\'e}},
  {Arsenijevic}, {Azaz}, {Balm}, {Beck}, {Bernstein}, {Bigot}, {Bijaoui},
  {Blasco}, {Bonfigli}, {Bono}, {Boudreault}, {Bressan}, {Brown}, {Brunet},
  {Bunclark}, {Buonanno}, {Butkevich}, {Carret}, {Carrion}, {Chemin},
  {Ch{\'e}reau}, {Corcione}, {Darmigny}, {de Boer}, {de Teodoro}, {de Zeeuw},
  {Delle Luche}, {Domingues}, {Dubath}, {Fodor}, {Fr{\'e}zouls}, {Fries},
  {Fustes}, {Fyfe}, {Gallardo}, {Gallegos}, {Gardiol}, {Gebran}, {Gomboc},
  {G{\'o}mez}, {Grux}, {Gueguen}, {Heyrovsky}, {Hoar}, {Iannicola}, {Isasi
  Parache}, {Janotto}, {Joliet}, {Jonckheere}, {Keil}, {Kim}, {Klagyivik},
  {Klar}, {Knude}, {Kochukhov}, {Kolka}, {Kos}, {Kutka}, {Lainey}, {LeBouquin},
  {Liu}, {Loreggia}, {Makarov}, {Marseille}, {Martayan}, {Martinez-Rubi},
  {Massart}, {Meynadier}, {Mignot}, {Munari}, {Nguyen}, {Nordlander}, {Ocvirk},
  {O'Flaherty}, {Olias Sanz}, {Ortiz}, {Osorio}, {Oszkiewicz}, {Ouzounis},
  {Palmer}, {Park}, {Pasquato}, {Peltzer}, {Peralta}, {P{\'e}turaud},
  {Pieniluoma}, {Pigozzi}, {Poels}, {Prat}, {Prod'homme}, {Raison}, {Rebordao},
  {Risquez}, {Rocca-Volmerange}, {Rosen}, {Ruiz-Fuertes}, {Russo}, {Sembay},
  {Serraller Vizcaino}, {Short}, {Siebert}, {Silva}, {Sinachopoulos}, {Slezak},
  {Soffel}, {Sosnowska}, {Strai{\v{z}}ys}, {ter Linden}, {Terrell}, {Theil},
  {Tiede}, {Troisi}, {Tsalmantza}, {Tur}, {Vaccari}, {Vachier}, {Valles}, {Van
  Hamme}, {Veltz}, {Virtanen}, {Wallut}, {Wichmann}, {Wilkinson}, {Ziaeepour},
  \& {Zschocke}}]{2016A&A...595A...2G}
{Gaia Collaboration}, {Brown}, A.~G.~A., {Vallenari}, A., {et~al.} 2016, \aap,
  595, A2

\bibitem[{{Gaia Collaboration} {et~al.}(2023){Gaia Collaboration}, {Vallenari},
  {Brown}, {Prusti}, {de Bruijne}, {Arenou}, {Babusiaux}, {Biermann},
  {Creevey}, {Ducourant}, {Evans}, {Eyer}, {Guerra}, {Hutton}, {Jordi},
  {Klioner}, {Lammers}, {Lindegren}, {Luri}, {Mignard}, {Panem}, {Pourbaix},
  {Randich}, {Sartoretti}, {Soubiran}, {Tanga}, {Walton}, {Bailer-Jones},
  {Bastian}, {Drimmel}, {Jansen}, {Katz}, {Lattanzi}, {van Leeuwen}, {Bakker},
  {Cacciari}, {Casta{\~n}eda}, {De Angeli}, {Fabricius}, {Fouesneau},
  {Fr{\'e}mat}, {Galluccio}, {Guerrier}, {Heiter}, {Masana}, {Messineo},
  {Mowlavi}, {Nicolas}, {Nienartowicz}, {Pailler}, {Panuzzo}, {Riclet}, {Roux},
  {Seabroke}, {Sordo}, {Th{\'e}venin}, {Gracia-Abril}, {Portell}, {Teyssier},
  {Altmann}, {Andrae}, {Audard}, {Bellas-Velidis}, {Benson}, {Berthier},
  {Blomme}, {Burgess}, {Busonero}, {Busso}, {C{\'a}novas}, {Carry}, {Cellino},
  {Cheek}, {Clementini}, {Damerdji}, {Davidson}, {de Teodoro}, {Nu{\~n}ez
  Campos}, {Delchambre}, {Dell'Oro}, {Esquej}, {Fern{\'a}ndez-Hern{\'a}ndez},
  {Fraile}, {Garabato}, {Garc{\'\i}a-Lario}, {Gosset}, {Haigron}, {Halbwachs},
  {Hambly}, {Harrison}, {Hern{\'a}ndez}, {Hestroffer}, {Hodgkin}, {Holl},
  {Jan{\ss}en}, {Jevardat de Fombelle}, {Jordan}, {Krone-Martins}, {Lanzafame},
  {L{\"o}ffler}, {Marchal}, {Marrese}, {Moitinho}, {Muinonen}, {Osborne},
  {Pancino}, {Pauwels}, {Recio-Blanco}, {Reyl{\'e}}, {Riello}, {Rimoldini},
  {Roegiers}, {Rybizki}, {Sarro}, {Siopis}, {Smith}, {Sozzetti}, {Utrilla},
  {van Leeuwen}, {Abbas}, {{\'A}brah{\'a}m}, {Abreu Aramburu}, {Aerts},
  {Aguado}, {Ajaj}, {Aldea-Montero}, {Altavilla}, {{\'A}lvarez}, {Alves},
  {Anders}, {Anderson}, {Anglada Varela}, {Antoja}, {Baines}, {Baker},
  {Balaguer-N{\'u}{\~n}ez}, {Balbinot}, {Balog}, {Barache}, {Barbato},
  {Barros}, {Barstow}, {Bartolom{\'e}}, {Bassilana}, {Bauchet}, {Becciani},
  {Bellazzini}, {Berihuete}, {Bernet}, {Bertone}, {Bianchi}, {Binnenfeld},
  {Blanco-Cuaresma}, {Blazere}, {Boch}, {Bombrun}, {Bossini}, {Bouquillon},
  {Bragaglia}, {Bramante}, {Breedt}, {Bressan}, {Brouillet}, {Brugaletta},
  {Bucciarelli}, {Burlacu}, {Butkevich}, {Buzzi}, {Caffau}, {Cancelliere},
  {Cantat-Gaudin}, {Carballo}, {Carlucci}, {Carnerero}, {Carrasco},
  {Casamiquela}, {Castellani}, {Castro-Ginard}, {Chaoul}, {Charlot}, {Chemin},
  {Chiaramida}, {Chiavassa}, {Chornay}, {Comoretto}, {Contursi}, {Cooper},
  {Cornez}, {Cowell}, {Crifo}, {Cropper}, {Crosta}, {Crowley}, {Dafonte},
  {Dapergolas}, {David}, {David}, {de Laverny}, {De Luise}, {De March}, {De
  Ridder}, {de Souza}, {de Torres}, {del Peloso}, {del Pozo}, {Delbo},
  {Delgado}, {Delisle}, {Demouchy}, {Dharmawardena}, {Di Matteo}, {Diakite},
  {Diener}, {Distefano}, {Dolding}, {Edvardsson}, {Enke}, {Fabre}, {Fabrizio},
  {Faigler}, {Fedorets}, {Fernique}, {Fienga}, {Figueras}, {Fournier},
  {Fouron}, {Fragkoudi}, {Gai}, {Garcia-Gutierrez}, {Garcia-Reinaldos},
  {Garc{\'\i}a-Torres}, {Garofalo}, {Gavel}, {Gavras}, {Gerlach}, {Geyer},
  {Giacobbe}, {Gilmore}, {Girona}, {Giuffrida}, {Gomel}, {Gomez},
  {Gonz{\'a}lez-N{\'u}{\~n}ez}, {Gonz{\'a}lez-Santamar{\'\i}a},
  {Gonz{\'a}lez-Vidal}, {Granvik}, {Guillout}, {Guiraud},
  {Guti{\'e}rrez-S{\'a}nchez}, {Guy}, {Hatzidimitriou}, {Hauser}, {Haywood},
  {Helmer}, {Helmi}, {Sarmiento}, {Hidalgo}, {Hilger}, {H{\l}adczuk}, {Hobbs},
  {Holland}, {Huckle}, {Jardine}, {Jasniewicz}, {Jean-Antoine Piccolo},
  {Jim{\'e}nez-Arranz}, {Jorissen}, {Juaristi Campillo}, {Julbe}, {Karbevska},
  {Kervella}, {Khanna}, {Kontizas}, {Kordopatis}, {Korn}, {K{\'o}sp{\'a}l},
  {Kostrzewa-Rutkowska}, {Kruszy{\'n}ska}, {Kun}, {Laizeau}, {Lambert},
  {Lanza}, {Lasne}, {Le Campion}, {Lebreton}, {Lebzelter}, {Leccia}, {Leclerc},
  {Lecoeur-Taibi}, {Liao}, {Licata}, {Lindstr{\o}m}, {Lister}, {Livanou},
  {Lobel}, {Lorca}, {Loup}, {Madrero Pardo}, {Magdaleno Romeo}, {Managau},
  {Mann}, {Manteiga}, {Marchant}, {Marconi}, {Marcos}, {Marcos Santos},
  {Mar{\'\i}n Pina}, {Marinoni}, {Marocco}, {Marshall}, {Martin Polo},
  {Mart{\'\i}n-Fleitas}, {Marton}, {Mary}, {Masip}, {Massari},
  {Mastrobuono-Battisti}, {Mazeh}, {McMillan}, {Messina}, {Michalik}, {Millar},
  {Mints}, {Molina}, {Molinaro}, {Moln{\'a}r}, {Monari}, {Mongui{\'o}},
  {Montegriffo}, {Montero}, {Mor}, {Mora}, {Morbidelli}, {Morel}, {Morris},
  {Muraveva}, {Murphy}, {Musella}, {Nagy}, {Noval}, {Oca{\~n}a}, {Ogden},
  {Ordenovic}, {Osinde}, {Pagani}, {Pagano}, {Palaversa}, {Palicio},
  {Pallas-Quintela}, {Panahi}, {Payne-Wardenaar}, {Pe{\~n}alosa Esteller},
  {Penttil{\"a}}, {Pichon}, {Piersimoni}, {Pineau}, {Plachy}, {Plum}, {Poggio},
  {Pr{\v{s}}a}, {Pulone}, {Racero}, {Ragaini}, {Rainer}, {Raiteri}, {Rambaux},
  {Ramos}, {Ramos-Lerate}, {Re Fiorentin}, {Regibo}, {Richards}, {Rios Diaz},
  {Ripepi}, {Riva}, {Rix}, {Rixon}, {Robichon}, {Robin}, {Robin}, {Roelens},
  {Rogues}, {Rohrbasser}, {Romero-G{\'o}mez}, {Rowell}, {Royer}, {Ruz Mieres},
  {Rybicki}, {Sadowski}, {S{\'a}ez N{\'u}{\~n}ez}, {Sagrist{\`a} Sell{\'e}s},
  {Sahlmann}, {Salguero}, {Samaras}, {Sanchez Gimenez}, {Sanna},
  {Santove{\~n}a}, {Sarasso}, {Schultheis}, {Sciacca}, {Segol}, {Segovia},
  {S{\'e}gransan}, {Semeux}, {Shahaf}, {Siddiqui}, {Siebert}, {Siltala},
  {Silvelo}, {Slezak}, {Slezak}, {Smart}, {Snaith}, {Solano}, {Solitro},
  {Souami}, {Souchay}, {Spagna}, {Spina}, {Spoto}, {Steele},
  {Steidelm{\"u}ller}, {Stephenson}, {S{\"u}veges}, {Surdej}, {Szabados},
  {Szegedi-Elek}, {Taris}, {Taylor}, {Teixeira}, {Tolomei}, {Tonello}, {Torra},
  {Torra}, {Torralba Elipe}, {Trabucchi}, {Tsounis}, {Turon}, {Ulla}, {Unger},
  {Vaillant}, {van Dillen}, {van Reeven}, {Vanel}, {Vecchiato}, {Viala},
  {Vicente}, {Voutsinas}, {Weiler}, {Wevers}, {Wyrzykowski}, {Yoldas}, {Yvard},
  {Zhao}, {Zorec}, {Zucker}, \& {Zwitter}}]{2023AA...674A...1G}
{Gaia Collaboration}, {Vallenari}, A., {Brown}, A.~G.~A., {et~al.} 2023, \aap,
  674, A1

\bibitem[{{Goldreich} \& {Soter}(1966)}]{1966Icar....5..375G}
{Goldreich}, P. \& {Soter}, S. 1966, \icarus, 5, 375

\bibitem[{{Gray}(2005)}]{2005PASP..117..711G}
{Gray}, D.~F. 2005, \pasp, 117, 711

\bibitem[{{G{\"u}nther} \& {Daylan}(2019)}]{allesfitter-code}
{G{\"u}nther}, M.~N. \& {Daylan}, T. 2019, {Allesfitter: Flexible Star and
  Exoplanet Inference From Photometry and Radial Velocity}, Astrophysics Source
  Code Library

\bibitem[{{G{\"u}nther} \& {Daylan}(2021)}]{allesfitter-paper}
{G{\"u}nther}, M.~N. \& {Daylan}, T. 2021, \apjs, 254, 13

\bibitem[{{Hatzes} \& {Rauer}(2015)}]{2015ApJ...810L..25H}
{Hatzes}, A.~P. \& {Rauer}, H. 2015, \apjl, 810, L25

\bibitem[{{Hord} {et~al.}(2021){Hord}, {Col{\'o}n}, {Kostov}, {Galgano},
  {Ricker}, {Vanderspek}, {Seager}, {Winn}, {Jenkins}, {Barclay}, {Caldwell},
  {Essack}, {Fausnaugh}, {Guerrero}, \& {Wohler}}]{2021AJ....162..263H}
{Hord}, B.~J., {Col{\'o}n}, K.~D., {Kostov}, V., {et~al.} 2021, \aj, 162, 263

\bibitem[{{Huang} \& {Lei}(2022)}]{2022AJ....164..232H}
{Huang}, X. \& {Lei}, H. 2022, \aj, 164, 232

\bibitem[{{Ito} \& {Ohtsuka}(2019)}]{2019MEEP....7....1I}
{Ito}, T. \& {Ohtsuka}, K. 2019, Monographs on Environment, Earth and Planets,
  7, 1

\bibitem[{{Kipping}(2013)}]{2013MNRAS.435.2152K}
{Kipping}, D.~M. 2013, \mnras, 435, 2152

\bibitem[{{Kopal}(1950)}]{1950HarCi.454....1K}
{Kopal}, Z. 1950, Harvard College Observatory Circular, 454, 1

\bibitem[{{Kozai}(1962)}]{1962AJ.....67..591K}
{Kozai}, Y. 1962, \aj, 67, 591

\bibitem[{{Laskar} \& {Robutel}(2001)}]{2001CeMDA..80...39L}
{Laskar}, J. \& {Robutel}, P. 2001, Celestial Mechanics and Dynamical
  Astronomy, 80, 39

\bibitem[{{Lee} \& {Peale}(2003)}]{2003ApJ...592.1201L}
{Lee}, M.~H. \& {Peale}, S.~J. 2003, \apj, 592, 1201

\bibitem[{{Lidov}(1962)}]{1962P&SS....9..719L}
{Lidov}, M.~L. 1962, \planss, 9, 719

\bibitem[{{Lightkurve Collaboration} {et~al.}(2018){Lightkurve Collaboration},
  {Cardoso}, {Hedges}, {Gully-Santiago}, {Saunders}, {Cody}, {Barclay}, {Hall},
  {Sagear}, {Turtelboom}, {Zhang}, {Tzanidakis}, {Mighell}, {Coughlin}, {Bell},
  {Berta-Thompson}, {Williams}, {Dotson}, \& {Barentsen}}]{2018ascl.soft12013L}
{Lightkurve Collaboration}, {Cardoso}, J. V. d.~M., {Hedges}, C., {et~al.}
  2018, {Lightkurve: Kepler and TESS time series analysis in Python}

\bibitem[{{Lindegren} {et~al.}(2021){Lindegren}, {Klioner}, {Hern{\'a}ndez},
  {Bombrun}, {Ramos-Lerate}, {Steidelm{\"u}ller}, {Bastian}, {Biermann}, {de
  Torres}, {Gerlach}, {Geyer}, {Hilger}, {Hobbs}, {Lammers}, {McMillan},
  {Stephenson}, {Casta{\~n}eda}, {Davidson}, {Fabricius}, {Gracia-Abril},
  {Portell}, {Rowell}, {Teyssier}, {Torra}, {Bartolom{\'e}}, {Clotet},
  {Garralda}, {Gonz{\'a}lez-Vidal}, {Torra}, {Abbas}, {Altmann}, {Anglada
  Varela}, {Balaguer-N{\'u}{\~n}ez}, {Balog}, {Barache}, {Becciani}, {Bernet},
  {Bertone}, {Bianchi}, {Bouquillon}, {Brown}, {Bucciarelli}, {Busonero},
  {Butkevich}, {Buzzi}, {Cancelliere}, {Carlucci}, {Charlot}, {Cioni},
  {Crosta}, {Crowley}, {del Peloso}, {del Pozo}, {Drimmel}, {Esquej}, {Fienga},
  {Fraile}, {Gai}, {Garcia-Reinaldos}, {Guerra}, {Hambly}, {Hauser},
  {Jan{\ss}en}, {Jordan}, {Kostrzewa-Rutkowska}, {Lattanzi}, {Liao}, {Licata},
  {Lister}, {L{\"o}ffler}, {Marchant}, {Masip}, {Mignard}, {Mints}, {Molina},
  {Mora}, {Morbidelli}, {Murphy}, {Pagani}, {Panuzzo}, {Pe{\~n}alosa Esteller},
  {Poggio}, {Re Fiorentin}, {Riva}, {Sagrist{\`a} Sell{\'e}s}, {Sanchez
  Gimenez}, {Sarasso}, {Sciacca}, {Siddiqui}, {Smart}, {Souami}, {Spagna},
  {Steele}, {Taris}, {Utrilla}, {van Reeven}, \&
  {Vecchiato}}]{2021A&A...649A...2L}
{Lindegren}, L., {Klioner}, S.~A., {Hern{\'a}ndez}, J., {et~al.} 2021, \aap,
  649, A2

\bibitem[{{Lovis} \& {Pepe}(2007)}]{2007A&A...468.1115L}
{Lovis}, C. \& {Pepe}, F. 2007, \aap, 468, 1115

\bibitem[{{Maciejewski}(2020)}]{2020AcA....70..181M}
{Maciejewski}, G. 2020, \actaa, 70, 181

\bibitem[{{Marcy} \& {Butler}(1992)}]{1992PASP..104..270M}
{Marcy}, G.~W. \& {Butler}, R.~P. 1992, \pasp, 104, 270

\bibitem[{{Mizuno}(1980)}]{1980PThPh..64..544M}
{Mizuno}, H. 1980, Progress of Theoretical Physics, 64, 544

\bibitem[{{Mustill} {et~al.}(2015){Mustill}, {Davies}, \&
  {Johansen}}]{2015ApJ...808...14M}
{Mustill}, A.~J., {Davies}, M.~B., \& {Johansen}, A. 2015, \apj, 808, 14

\bibitem[{{Naoz}(2016)}]{2016ARA&A..54..441N}
{Naoz}, S. 2016, \araa, 54, 441

\bibitem[{{Naoz} {et~al.}(2011){Naoz}, {Farr}, {Lithwick}, {Rasio}, \&
  {Teyssandier}}]{2011Natur.473..187N}
{Naoz}, S., {Farr}, W.~M., {Lithwick}, Y., {Rasio}, F.~A., \& {Teyssandier}, J.
  2011, \nat, 473, 187

\bibitem[{{Nelson} {et~al.}(2017){Nelson}, {Ford}, \&
  {Rasio}}]{2017AJ....154..106N}
{Nelson}, B.~E., {Ford}, E.~B., \& {Rasio}, F.~A. 2017, \aj, 154, 106

\bibitem[{{Ni}(2018)}]{2018A&A...613A..32N}
{Ni}, D. 2018, \aap, 613, A32

\bibitem[{{Niedzielski} {et~al.}(2007){Niedzielski}, {Konacki}, {Wolszczan},
  {Nowak}, {Maciejewski}, {Gelino}, {Shao}, {Shetrone}, \&
  {Ramsey}}]{2007ApJ...669.1354N}
{Niedzielski}, A., {Konacki}, M., {Wolszczan}, A., {et~al.} 2007, \apj, 669,
  1354

\bibitem[{{Niedzielski} {et~al.}(2015){Niedzielski}, {Villaver}, {Wolszczan},
  {Adam{\'o}w}, {Kowalik}, {Maciejewski}, {Nowak}, {Garc{\'\i}a-Hern{\'a}ndez},
  {Deka}, \& {Adamczyk}}]{2015A&A...573A..36N}
{Niedzielski}, A., {Villaver}, E., {Wolszczan}, A., {et~al.} 2015, \aap, 573,
  A36

\bibitem[{{Nobili} \& {Roxburgh}(1986)}]{1986IAUS..114..105N}
{Nobili}, A. \& {Roxburgh}, I.~W. 1986, in Relativity in Celestial Mechanics
  and Astrometry. High Precision Dynamical Theories and Observational
  Verifications, ed. J.~{Kovalevsky} \& V.~A. {Brumberg}, Vol. 114, 105

\bibitem[{{Nowak}(2012)}]{2012Nowak}
{Nowak}, G. 2012, PhD thesis, Nicolaus Copernicus University

\bibitem[{{Nowak} {et~al.}(2013){Nowak}, {Niedzielski}, {Wolszczan},
  {Adam{\'o}w}, \& {Maciejewski}}]{2013ApJ...770...53N}
{Nowak}, G., {Niedzielski}, A., {Wolszczan}, A., {Adam{\'o}w}, M., \&
  {Maciejewski}, G. 2013, \apj, 770, 53

\bibitem[{{Pepe} {et~al.}(2002){Pepe}, {Mayor}, {Galland}, {Naef}, {Queloz},
  {Santos}, {Udry}, \& {Burnet}}]{2002A&A...388..632P}
{Pepe}, F., {Mayor}, M., {Galland}, F., {et~al.} 2002, \aap, 388, 632

\bibitem[{{Pepper} {et~al.}(2020){Pepper}, {Kane}, {Rodriguez}, {Hinkel},
  {Eastman}, {Daylan}, {Mocnik}, {Dalba}, {Gaudi}, {Fetherolf}, {Stassun},
  {Campante}, {Vanderburg}, {Huber}, {Bossini}, {Crossfield}, {Howell},
  {Stephens}, {Furlan}, {Ricker}, {Vanderspek}, {Latham}, {Seager}, {Winn},
  {Jenkins}, {Twicken}, {Rose}, {Smith}, {Glidden}, {Levine}, {Rinehart},
  {Collins}, {Mann}, {Burt}, {James}, {Siverd}, \&
  {G{\"u}nther}}]{2020AJ....159..243P}
{Pepper}, J., {Kane}, S.~R., {Rodriguez}, J.~E., {et~al.} 2020, \aj, 159, 243

\bibitem[{{Petrovich}(2015)}]{2015ApJ...805...75P}
{Petrovich}, C. 2015, \apj, 805, 75

\bibitem[{{Pollack} {et~al.}(1996){Pollack}, {Hubickyj}, {Bodenheimer},
  {Lissauer}, {Podolak}, \& {Greenzweig}}]{1996Icar..124...62P}
{Pollack}, J.~B., {Hubickyj}, O., {Bodenheimer}, P., {et~al.} 1996, \icarus,
  124, 62

\bibitem[{{Ragozzine} \& {Wolf}(2009)}]{2009ApJ...698.1778R}
{Ragozzine}, D. \& {Wolf}, A.~S. 2009, \apj, 698, 1778

\bibitem[{{Ramsey} {et~al.}(1998){Ramsey}, {Adams}, {Barnes}, {Booth},
  {Cornell}, {Fowler}, {Gaffney}, {Glaspey}, {Good}, {Hill}, {Kelton},
  {Krabbendam}, {Long}, {MacQueen}, {Ray}, {Ricklefs}, {Sage}, {Sebring},
  {Spiesman}, \& {Steiner}}]{1998SPIE.3352...34R}
{Ramsey}, L.~W., {Adams}, M.~T., {Barnes}, T.~G., {et~al.} 1998, in Society of
  Photo-Optical Instrumentation Engineers (SPIE) Conference Series, Vol. 3352,
  Advanced Technology Optical/IR Telescopes VI, ed. L.~M. {Stepp}, 34--42

\bibitem[{{Rauer} {et~al.}(2014){Rauer}, {Catala}, {Aerts}, {Appourchaux},
  {Benz}, {Brandeker}, {Christensen-Dalsgaard}, {Deleuil}, {Gizon}, {Goupil},
  {G{\"u}del}, {Janot-Pacheco}, {Mas-Hesse}, {Pagano}, {Piotto}, {Pollacco},
  {Santos}, {Smith}, {Su{\'a}rez}, {Szab{\'o}}, {Udry}, {Adibekyan}, {Alibert},
  {Almenara}, {Amaro-Seoane}, {Eiff}, {Asplund}, {Antonello}, {Barnes},
  {Baudin}, {Belkacem}, {Bergemann}, {Bihain}, {Birch}, {Bonfils}, {Boisse},
  {Bonomo}, {Borsa}, {Brand{\~a}o}, {Brocato}, {Brun}, {Burleigh}, {Burston},
  {Cabrera}, {Cassisi}, {Chaplin}, {Charpinet}, {Chiappini}, {Church},
  {Csizmadia}, {Cunha}, {Damasso}, {Davies}, {Deeg}, {D{\'\i}az}, {Dreizler},
  {Dreyer}, {Eggenberger}, {Ehrenreich}, {Eigm{\"u}ller}, {Erikson}, {Farmer},
  {Feltzing}, {de Oliveira Fialho}, {Figueira}, {Forveille}, {Fridlund},
  {Garc{\'\i}a}, {Giommi}, {Giuffrida}, {Godolt}, {Gomes da Silva}, {Granzer},
  {Grenfell}, {Grotsch-Noels}, {G{\"u}nther}, {Haswell}, {Hatzes},
  {H{\'e}brard}, {Hekker}, {Helled}, {Heng}, {Jenkins}, {Johansen},
  {Khodachenko}, {Kislyakova}, {Kley}, {Kolb}, {Krivova}, {Kupka}, {Lammer},
  {Lanza}, {Lebreton}, {Magrin}, {Marcos-Arenal}, {Marrese}, {Marques},
  {Martins}, {Mathis}, {Mathur}, {Messina}, {Miglio}, {Montalban}, {Montalto},
  {Monteiro}, {Moradi}, {Moravveji}, {Mordasini}, {Morel}, {Mortier},
  {Nascimbeni}, {Nelson}, {Nielsen}, {Noack}, {Norton}, {Ofir}, {Oshagh},
  {Ouazzani}, {P{\'a}pics}, {Parro}, {Petit}, {Plez}, {Poretti}, {Quirrenbach},
  {Ragazzoni}, {Raimondo}, {Rainer}, {Reese}, {Redmer}, {Reffert},
  {Rojas-Ayala}, {Roxburgh}, {Salmon}, {Santerne}, {Schneider}, {Schou},
  {Schuh}, {Schunker}, {Silva-Valio}, {Silvotti}, {Skillen}, {Snellen}, {Sohl},
  {Sousa}, {Sozzetti}, {Stello}, {Strassmeier}, {{\v{S}}vanda}, {Szab{\'o}},
  {Tkachenko}, {Valencia}, {Van Grootel}, {Vauclair}, {Ventura}, {Wagner},
  {Walton}, {Weingrill}, {Werner}, {Wheatley}, \&
  {Zwintz}}]{2014ExA....38..249R}
{Rauer}, H., {Catala}, C., {Aerts}, C., {et~al.} 2014, Experimental Astronomy,
  38, 249

\bibitem[{{Rein} \& {Liu}(2012)}]{2012A&A...537A.128R}
{Rein}, H. \& {Liu}, S.~F. 2012, \aap, 537, A128

\bibitem[{{Rein} \& {Spiegel}(2015)}]{2015MNRAS.446.1424R}
{Rein}, H. \& {Spiegel}, D.~S. 2015, \mnras, 446, 1424

\bibitem[{{Rein} \& {Tamayo}(2015)}]{2015MNRAS.452..376R}
{Rein}, H. \& {Tamayo}, D. 2015, \mnras, 452, 376

\bibitem[{{Rey} {et~al.}(2018){Rey}, {Bouchy}, {Stalport}, {Deleuil},
  {H{\'e}brard}, {Almenara}, {Alonso}, {Barros}, {Bonomo}, {Cazalet},
  {Delisle}, {D{\'\i}az}, {Fridlund}, {Guenther}, {Guillot}, {Montagnier},
  {Moutou}, {Lovis}, {Queloz}, {Santerne}, \& {Udry}}]{2018A&A...619A.115R}
{Rey}, J., {Bouchy}, F., {Stalport}, M., {et~al.} 2018, \aap, 619, A115

\bibitem[{{Ricker} {et~al.}(2015){Ricker}, {Winn}, {Vanderspek}, {Latham},
  {Bakos}, {Bean}, {Berta-Thompson}, {Brown}, {Buchhave}, {Butler}, {Butler},
  {Chaplin}, {Charbonneau}, {Christensen-Dalsgaard}, {Clampin}, {Deming},
  {Doty}, {De Lee}, {Dressing}, {Dunham}, {Endl}, {Fressin}, {Ge}, {Henning},
  {Holman}, {Howard}, {Ida}, {Jenkins}, {Jernigan}, {Johnson}, {Kaltenegger},
  {Kawai}, {Kjeldsen}, {Laughlin}, {Levine}, {Lin}, {Lissauer}, {MacQueen},
  {Marcy}, {McCullough}, {Morton}, {Narita}, {Paegert}, {Palle}, {Pepe},
  {Pepper}, {Quirrenbach}, {Rinehart}, {Sasselov}, {Sato}, {Seager},
  {Sozzetti}, {Stassun}, {Sullivan}, {Szentgyorgyi}, {Torres}, {Udry}, \&
  {Villasenor}}]{2015JATIS...1a4003R}
{Ricker}, G.~R., {Winn}, J.~N., {Vanderspek}, R., {et~al.} 2015, Journal of
  Astronomical Telescopes, Instruments, and Systems, 1, 014003

\bibitem[{{Savitzky} \& {Golay}(1964)}]{1964AnaCh..36.1627S}
{Savitzky}, A. \& {Golay}, M.~J.~E. 1964, Analytical Chemistry, 36, 1627

\bibitem[{{Schlaufman}(2018)}]{2018ApJ...853...37S}
{Schlaufman}, K.~C. 2018, \apj, 853, 37

\bibitem[{{Schwarzenberg-Czerny}(1989)}]{1989MNRAS.241..153S}
{Schwarzenberg-Czerny}, A. 1989, \mnras, 241, 153

\bibitem[{{Schwarzenberg-Czerny} \& {Beaulieu}(2006)}]{2006MNRAS.365..165S}
{Schwarzenberg-Czerny}, A. \& {Beaulieu}, J.~P. 2006, \mnras, 365, 165

\bibitem[{{Sha} {et~al.}(2023){Sha}, {Vanderburg}, {Huang}, {Armstrong},
  {Brahm}, {Giacalone}, {Wood}, {Collins}, {Nielsen}, {Hobson}, {Ziegler},
  {Howell}, {Torres-Miranda}, {Mann}, {Zhou}, {Delgado-Mena}, {Rojas}, {Abe},
  {Trifonov}, {Adibekyan}, {Sousa}, {Fajardo-Acosta}, {Guillot}, {Howard},
  {Littlefield}, {Hawthorn}, {Schmider}, {Eberhardt}, {Tan}, {Osborn},
  {Schwarz}, {Str{\o}m}, {Jord{\'a}n}, {Wang}, {Henning}, {Massey}, {Law},
  {Stockdale}, {Furlan}, {Srdoc}, {Wheatley}, {Barrado Navascu{\'e}s},
  {Lissauer}, {Stassun}, {Ricker}, {Vanderspek}, {Latham}, {Winn}, {Seager},
  {Jenkins}, {Barclay}, {Bouma}, {Christiansen}, {Guerrero}, \&
  {Rose}}]{2023MNRAS.524.1113S}
{Sha}, L., {Vanderburg}, A.~M., {Huang}, C.~X., {et~al.} 2023, \mnras, 524,
  1113

\bibitem[{{Sousa} {et~al.}(2021){Sousa}, {Adibekyan}, {Delgado-Mena}, {Santos},
  {Rojas-Ayala}, {Soares}, {Legoinha}, {Ulmer-Moll}, {Camacho}, {Barros},
  {Demangeon}, {Hoyer}, {Israelian}, {Mortier}, {Tsantaki}, \&
  {Monteiro}}]{2021AA...656A..53S}
{Sousa}, S.~G., {Adibekyan}, V., {Delgado-Mena}, E., {et~al.} 2021, \aap, 656,
  A53

\bibitem[{{Spiegel} {et~al.}(2011){Spiegel}, {Burrows}, \&
  {Milsom}}]{2011ApJ...727...57S}
{Spiegel}, D.~S., {Burrows}, A., \& {Milsom}, J.~A. 2011, \apj, 727, 57

\bibitem[{{Takeda} {et~al.}(2005){Takeda}, {Ohkubo}, {Sato}, {Kambe}, \&
  {Sadakane}}]{2005PASJ...57...27T}
{Takeda}, Y., {Ohkubo}, M., {Sato}, B., {Kambe}, E., \& {Sadakane}, K. 2005,
  \pasj, 57, 27

\bibitem[{{Tamayo} {et~al.}(2020){Tamayo}, {Rein}, {Shi}, \&
  {Hernandez}}]{2020MNRAS.491.2885T}
{Tamayo}, D., {Rein}, H., {Shi}, P., \& {Hernandez}, D.~M. 2020, \mnras, 491,
  2885

\bibitem[{{Tull}(1998)}]{1998SPIE.3355..387T}
{Tull}, R.~G. 1998, in Society of Photo-Optical Instrumentation Engineers
  (SPIE) Conference Series, Vol. 3355, Optical Astronomical Instrumentation,
  ed. S.~{D'Odorico}, 387--398

\bibitem[{{Udry}(2010)}]{2010lyot.confE..11U}
{Udry}, S. 2010, in In the Spirit of Lyot 2010, ed. A.~{Boccaletti}, E11

\bibitem[{{Valenti} \& {Piskunov}(1996)}]{1996A&AS..118..595V}
{Valenti}, J.~A. \& {Piskunov}, N. 1996, \aaps, 118, 595

\bibitem[{{van Leeuwen}(2007)}]{2007A&A...474..653V}
{van Leeuwen}, F. 2007, \aap, 474, 653

\bibitem[{{Volpi} \& {Libert}(2024)}]{2024A&A...683A.193V}
{Volpi}, M. \& {Libert}, A.-S. 2024, \aap, 683, A193

\bibitem[{{von Zeipel}(1910)}]{1910AN....183..345V}
{von Zeipel}, H. 1910, Astronomische Nachrichten, 183, 345

\bibitem[{{Winn} \& {Fabrycky}(2015)}]{2015ARA&A..53..409W}
{Winn}, J.~N. \& {Fabrycky}, D.~C. 2015, \araa, 53, 409

\bibitem[{{Wisdom} \& {Holman}(1991)}]{1991AJ....102.1528W}
{Wisdom}, J. \& {Holman}, M. 1991, \aj, 102, 1528

\bibitem[{{Wu} \& {Murray}(2003)}]{2003ApJ...589..605W}
{Wu}, Y. \& {Murray}, N. 2003, \apj, 589, 605

\bibitem[{{Yoder}(1979)}]{1979Natur.279..767Y}
{Yoder}, C.~F. 1979, \nat, 279, 767

\end{thebibliography}

\appendix

\section{Supplementary materials}

Tables~\ref{tab.HET} and \ref{tab.HARPSn} present the details of individual RV observations from HRS/HET and HIRES-N/TGN, respectively. Table~\ref{tab.systemicParams} provides the systemic parameters derived in this study.

\begin{table}[h!]
\caption{Individual Doppler observations from HRS/HET.} 
\label{tab.HET}      
\centering                  
\begin{tabular}{l c c c}      
\hline
\hline
$\rm{BJD}_{\rm{TDB}}$ & RV            & $\sigma_{\rm{RV}}$ & BS \\
                      & (km s$^{-1}$) & (km s$^{-1}$)      & (m s$^{-1}$) \\
 \hline
\hline
$2453754.931620$ &  $0.07698$ & $0.00955$ &  $-4.26$ \\
$2453775.866019$ & $-0.11714$ & $0.00910$ & $-15.92$ \\
$2453800.785891$ & $-0.03834$ & $0.00686$ &  $31.47$ \\
$2453835.879850$ & $-0.26580$ & $0.00960$ & $-11.06$ \\
$2453889.752627$ &  $0.11391$ & $0.00713$ & $-47.41$ \\
$2453910.700590$ & $-0.10492$ & $0.00843$ & $-73.12$ \\
$2453937.627106$ &  $0.19614$ & $0.00615$ &   $3.78$ \\
$2454155.829329$ & $-0.17799$ & $0.00752$ & $-11.29$ \\
$2454161.806250$ & $-0.17900$ & $0.00834$ &   $0.34$ \\
$2454162.013218$ & $-0.13211$ & $0.00733$ &   $6.76$ \\
$2454167.789387$ & $-0.20552$ & $0.00647$ &  $-5.87$ \\
$2454173.769005$ & $-0.25126$ & $0.00858$ &  $-8.40$ \\
$2454174.757523$ & $-0.03856$ & $0.00760$ & $-37.54$ \\
$2454180.750972$ & $-0.05006$ & $0.00777$ & $-39.03$ \\
$2454186.734051$ & $-0.07518$ & $0.00749$ & $-19.01$ \\
$2454216.654954$ & $-0.25227$ & $0.00577$ & $-40.67$ \\
$2454221.651389$ &  $0.04484$ & $0.00735$ & $-83.02$ \\
$2454247.779954$ & $-0.17380$ & $0.00606$ &  $22.09$ \\
$2454255.766701$ &  $0.11980$ & $0.00822$ & $-31.83$ \\
$2454262.733357$ &  $0.20291$ & $0.01034$ &   $4.43$ \\
$2454262.741482$ &  $0.21158$ & $0.00847$ &  $18.72$ \\
$2454300.628061$ &  $0.19736$ & $0.00944$ & $-50.03$ \\
$2454462.987002$ & $-0.04824$ & $0.00828$ &  $-6.80$ \\
$2455014.680474$ & $-0.13054$ & $0.01213$ &   $3.42$ \\
$2455040.626748$ &  $0.12471$ & $0.00854$ & $-29.54$ \\
$2455314.850961$ & $-0.21131$ & $0.00864$ &   $5.36$ \\
$2455555.005828$ & $-0.04931$ & $0.01287$ &  $12.42$ \\
$2455556.997251$ &  $0.23908$ & $0.01042$ &  $30.85$ \\
$2455557.994444$ &  $0.25211$ & $0.01487$ & $-73.31$ \\
$2455563.981482$ &  $0.17646$ & $0.01099$ & $-22.05$ \\
$2455567.970249$ &  $0.09616$ & $0.01237$ &  $-8.12$ \\
$2455629.997095$ &  $0.17312$ & $0.01110$ &   $6.23$ \\
$2455630.789641$ &  $0.20592$ & $0.00637$ &   $1.89$ \\
$2455634.966296$ &  $0.03518$ & $0.00753$ & $-66.87$ \\
$2455637.772726$ &  $0.19955$ & $0.00696$ & $-35.02$ \\
$2455639.968102$ & $-0.19939$ & $0.00825$ &  $18.20$ \\
$2455640.770966$ & $-0.06143$ & $0.00795$ & $-31.27$ \\
$2455703.775810$ &  $0.15091$ & $0.00967$ & $-44.06$ \\
$2455729.691742$ &  $0.24031$ & $0.00979$ & $-18.91$ \\
$2455734.701291$ &  $0.19321$ & $0.00918$ &  $-7.25$ \\
$2455746.656314$ &  $0.19373$ & $0.00989$ & $-47.51$ \\
$2455751.636308$ &  $0.03605$ & $0.00962$ & $-30.54$ \\
$2455933.974525$ & $-0.17229$ & $0.00752$ &  $18.99$ \\
$2455948.888773$ &  $0.16626$ & $0.00731$ &  $-0.40$ \\
$2455964.868542$ & $-0.23200$ & $0.00740$ & $-45.38$ \\
$2455978.820943$ &  $0.04186$ & $0.00704$ & $-16.13$ \\
$2456003.963964$ &  $0.12330$ & $0.00726$ & $-17.41$ \\
$2456043.643206$ &  $0.07747$ & $0.00752$ & $-11.93$ \\
$2456073.755683$ &  $0.15241$ & $0.00702$ & $-31.82$ \\
$2456398.679884$ &  $0.14608$ & $0.00892$ &  $26.54$ \\
$2456461.714097$ & $-0.24824$ & $0.00839$ & $-15.89$ \\
\hline                                   
\end{tabular}
\tablefoot{$\rm{BJD}_{\rm{TDB}}$ is given for the exposure centroid. RV and $\sigma_{\rm{RV}}$ are the determined values of radial velocity and its uncertainty, respectively. BS is the value of the bisector span. All values are intentionally left unrounded.}
\end{table}

\begin{table}[h]
\caption{Individual Doppler observations from HIRES-N/TNG.} 
\label{tab.HARPSn}      
\centering                  
\begin{tabular}{l c c c}      
\hline
\hline
$\rm{BJD}_{\rm{TDB}}$ & RV            & $\sigma_{\rm{RV}}$ & BS \\
                      & (km s$^{-1}$) & (km s$^{-1}$)      & (m s$^{-1}$) \\
 \hline
\noalign{\smallskip}
$2456277.776746$ & $-29.41166$ & $0.00207$ & $45.68$\\
$2456294.750929$ & $-29.04077$ & $0.00131$ & $61.11$\\
$2456411.479219$ & $-29.08413$ & $0.00161$ & $59.89$\\
$2456470.431029$ & $-29.05977$ & $0.00197$ & $40.06$\\
$2456647.758777$ & $-29.13123$ & $0.00273$ & $47.87$\\
$2456685.688117$ & $-29.03608$ & $0.00283$ & $42.82$\\
$2456740.588126$ & $-29.05898$ & $0.00280$ & $28.32$\\
$2456770.557675$ & $-29.14227$ & $0.00198$ & $50.07$\\
$2456795.507729$ & $-29.10550$ & $0.00169$ & $46.50$\\
$2456837.449723$ & $-29.25237$ & $0.00166$ & $42.86$\\
$2457035.768792$ & $-29.09250$ & $0.00253$ & $54.72$\\
$2457066.729477$ & $-29.08357$ & $0.00270$ & $37.56$\\
$2457135.516109$ & $-29.20869$ & $0.00176$ & $60.33$\\
$2457168.476646$ & $-29.36729$ & $0.00089$ & $51.64$\\
$2457196.430544$ & $-29.15954$ & $0.00131$ & $50.08$\\
$2457196.524173$ & $-29.17091$ & $0.00133$ & $52.21$\\
$2457238.372546$ & $-29.12011$ & $0.00153$ & $55.61$\\
$2460023.765436$ & $-28.92247$ & $0.00170$ & $40.87$\\
\hline                                   
\end{tabular}
\tablefoot{Designations as in Table~\ref{tab.HET}.}
\end{table}

\begin{table*}[h]
\caption{Systemic parameters for HD~118203 from the two-planet model (Sect.~\ref{SSect:Results.2plSystem}) and the astrometric constraints (Sect.~\ref{SSect:Results.Astrometry}).} 
\label{tab.systemicParams}      
\centering                  
\begin{tabular}{l l c c c c}      
\hline\hline                
 & Parameter         & Prior  & This paper & Pepper & Castro-Gonz\'alez \\
 &                   &        &            & et al. (2020) & et al. (2024)\\
\hline
\multicolumn{6}{l}{HD~118203~b}\\
& Orbital period, $P_{\rm{orb,b}}$ (d) & $\mathcal{U}(6.134,6.136)$ & $6.1349890(13)$ & $6.134985^{+(29)}_{-(30)}$ & 6.1349847(20)\\
& Conjunction time, $T_{\rm{0,b}}$ ($\mathrm{BJD_{TDB}}$)$^{\rm {a}}$ & $\mathcal{U}(8712.56,8712.76)$ & $8712.66178^{+(20)}_{-(19)}$ & $8712.66175(17)$ & $...$ \\
& RV amplitude, $K_{\rm{b}}$ $(\rm{km~s}^{-1})$ & $\mathcal{U}(0.15,0.29)$ & $0.2222\pm0.0021$ & $0.2180^{+0.0044}_{-0.0043}$ & $0.2180^{+0.0040}_{-0.0039}$\\
& Orbital eccentricity, $e_{\rm{b}}$ & $\mathcal{U}(0,1)$ & $0.301 \pm 0.006$ & $0.314 \pm 0.017$ & $0.316 \pm 0.020$ \\
& Argument of periastron, $\omega_{\rm{b}}$ $(\degr)$ & $\mathcal{U}(0,360)$ & $157.4 \pm 1.9$ & $152.8 \pm 3.1$ & $152.5 \pm 2.7$ \\
& Radius ratio, $R_{\rm{b}}/R_{\star}$ & $\mathcal{U}(0.05,0.06)$ & $0.05541_{-0.00017}^{+0.00025}$ & $0.05552^{+0.00019}_{-0.00017}$ & $0.05546^{+0.00024}_{-0.00022}$\\
& Scaled sum of radii, $(R_{\rm{b}}+R_{\star})/a_{\rm b}$ & $\mathcal{U}(0.1,0.2)$ & $0.1415_{-0.0020}^{+0.0026}$ & $...$  & $...$\\
& Orbital inclination, $i_{\rm{b}}$ $(\degr)$ & $\mathcal{U}(80,90)$ & $88.9_{-1.0}^{+0.8}$ & $88.88^{+0.77}_{-0.97}$ & $89.16^{+0.57}_{-0.65}$\\
& Transit impact parameter, $b_{\rm{b}}$ & $...$ & $0.12_{-0.08}^{+0.10}$ & $0.111^{+0.095}_{-0.076}$ & $0.14\pm0.10$\\
& Transit depth, $\delta_\mathrm{tr,b}$ (ppth) & $...$ & $3.564_{-0.024}^{+0.023}$ & $3.083_{-0.018}^{+0.021}$ & $3.075 \pm 0.025$\\ 
& Total transit duration, $\tau_{\rm{14,b}}$ (h) & $...$ & $5.638_{-0.011}^{+0.013}$ & $5.650^{+0.014}_{-0.012}$ & $...$\\
& Scaled semi-major axis, $a_{\rm b}/R_{\star}$ & $...$ & $7.46_{-0.14}^{+0.11}$ & $7.23^{+0.13}_{-0.14}$ & $9.32 \pm 0.31$\\
& Semi-major axis, $a_{\rm{b}}$ (au) & $...$ & $0.0701\pm0.0004$ & $0.0707^{+0.0010}_{-0.0011}$ & $0.08635 \pm 0.00057$\\
& Mass, $M_{\rm{b}}$ $(M_{\rm{Jup}})$ & $...$ & $2.182\pm0.033$ & $2.166^{+0.074}_{-0.079}$ & $2.282 \pm 0.045$\\
& Radius, $R_{\rm{b}}$ $(R_{\rm{Jup}})$ & $...$ & $1.12\pm0.09$ & $1.136^{+0.029}_{-0.028}$ & $1.076 \pm 0.035$\\
& Density, $\rho_{\rm{b}}$ $(\rm{g~cm}^{-3})$ & $...$ & $2.1\pm0.5$ & $1.83 \pm 0.12$ & $2.27 \pm 0.23$\\
& Surface gravity, $g_\mathrm{b}$ $(\rm{m~s}^{-2})$ & $...$ & $45.6_{-1.9}^{+1.5}$ & $41.6 \pm 1.9$ & $48.9 \pm 3.4$\\ 
& Equilibrium temperature$^{\rm {b}}$, $T_\mathrm{eq,b}$ (K) & $...$ & $1371_{-17}^{+16}$ & $1367\pm23$ & $1360 \pm 23$\\ 
\multicolumn{6}{l}{HD~118203~c} \\
& Orbital period, $P_{\rm{orb,c}}$ (d) & $\mathcal{U}(3000,10000)$ & $5070_{-230}^{+240}$ & $...$ & $...$ \\
& Conjunction time, $T_{\rm{0,c}}$ ($\mathrm{BJD_{TDB}}$)$^{\rm {a}}$ & $\mathcal{U}(4000,10000)$ & $6825_{-63}^{+68}$ & $...$ & $...$ \\
& RV amplitude, $K_{\rm{c}}$ $(\rm{km~s}^{-1})$ & $\mathcal{U}(0.01,0.4)$ & $0.118_{-0.010}^{+0.011}$ & $...$ & $...$ \\
& Orbital eccentricity, $e_{\rm{c}}$ & $\mathcal{U}(0,1)$ & $0.257\pm0.034$ & $...$ & $...$ \\
& Argument of periastron, $\omega_{\rm{c}}$ $(\degr)$ & $\mathcal{U}(0,360)$ & $175.4_{-4.6}^{+4.3}$ & $...$ & $...$  \\
& Orbital inclination$^{\rm {c}}$, $i_{\rm{c}}$ $(\degr)$ & $\mathcal{U}(0,180)$ & $95^{+15}_{-19}$ & $...$ & $...$ \\
& Semi-major axis, $a_{\rm{c}}$ (au) & $...$ & $6.20 \pm 0.20$ & $...$ & $...$ \\
& Mass, $M_{\rm{c}}$ $(M_{\rm{Jup}})$ & $...$ & $11.1_{-1.0}^{+1.3}$ & $...$ & $...$ \\
& Equilibrium temperature$^{\rm {b}}$, $T_\mathrm{eq,c}$ (K) & $...$ & $146 \pm 3$ & $...$ \\ 
\multicolumn{6}{l}{Common} \\
& $q_1$ LD coefficient, $q_{\rm 1,TESS}$ & $\mathcal{U}(0,1)$ & $0.237_{-0.034}^{+0.039}$ & $...$ & $...$ \\
& $q_2$ LD coefficient, $q_{\rm 2,TESS}$ & $\mathcal{U}(0,1)$ & $0.369_{-0.079}^{+0.077}$ & $...$ & $...$ \\
& Linear LD coefficient, $u_{\rm 1,TESS}$ & $...$ & $0.36 \pm 0.05$ & $0.355\pm0.063$ & $0.347^{+0.041}_{-0.042}$ \\
& Quadratic LD coefficient, $u_{\rm 2,TESS}$ & $...$ & $0.13 \pm 0.08$ & $0.14\pm0.10$ & $0.150^{+0.071}_{-0.069}$ \\
& Stellar density from transits, $\rho_{\rm \star,tr}$ ($\rho_{\odot}$) & $...$ & $0.1482_{-0.0080}^{+0.0063}$ & $...$ & $...$ \\ 
& Stellar radius$^{\rm {d}}$, $R_{\rm \star,tr}$ ($R_{\odot}$) & $...$ & $2.019_{-0.031}^{+0.038}$ & $...$ & $...$ \\ 
& Barycentric $\rm{RV_{ELODIE}}$ $(\rm{km~s}^{-1})$ & $\mathcal{U}(-30.0, -28.5)$ & $-29.288\pm0.011$ & $-29.312 \pm 0.003$ & $-29.3643 \pm 0.0045$\\
& $\rm{RV_{HRS}}$ offset $(\rm{km~s}^{-1})$ & $\mathcal{U}(-0.5, 0.5)$ & $-0.020 \pm 0.009$ & $...$ & $...$ \\
& Barycentric $\rm{RV_{HARPS-N}}$ $(\rm{km~s}^{-1})$ & $\mathcal{U}(-30.0, -28.5)$ & $-29.177 \pm 0.008$ & $...$ & $...$ \\
& RV jitter $\ln{\sigma_\mathrm{jitter, ELODIE}}$ $(\ln{ \mathrm{km\,s^{-1}} })$ & $\mathcal{U}(-15, 0)$ & $-4.29_{-0.19}^{+0.18}$ & $-4.19_{-0.20}^{+0.19}$ & $-4.23_{-0.19}^{+0.18}$\\ 
& RV jitter $\ln{\sigma_\mathrm{jitter, HRS}}$ $(\ln{ \mathrm{km\,s^{-1}} })$ & $\mathcal{U}(-15, 0)$ & $-4.20\pm0.14$ & $...$ & $...$ \\ 
& RV jitter $\ln{\sigma_\mathrm{jitter, HARPS-N}}$ $(\ln{ \mathrm{km\,s^{-1}} })$ & $\mathcal{U}(-15, 0)$ & $-4.91_{-0.25}^{+0.26}$ & $...$ & $...$ \\ 
& RV jitter $\sigma_\mathrm{jitter, ELODIE}$ $(\mathrm{km\,s^{-1}})$ & $...$ & $0.0137_{-0.0027}^{+0.0024}$ & $0.0152_{-0.0027}^{+0.0031}$ & $0.0145_{-0.0025}^{+0.0028}$\\ 
& RV jitter $\sigma_\mathrm{jitter, HRS}$ $(\mathrm{km\,s^{-1}})$ & $...$ & $0.0150 \pm 0.0021$ & $...$ & $...$ \\ 
& RV jitter $\sigma_\mathrm{jitter, HARPS-N}$ $(\mathrm{km\,s^{-1}})$ & $...$ & $0.0074_{-0.0018}^{+0.0019}$ & $...$ & $...$ \\ 
\hline                                   
\end{tabular}
\tablefoot{Priors are given for free parameters of the fit. The derived quantities are those without priors. $^{\rm {a}}$ Given as $\mathrm{BJD_{TDB}}-2450000$. $^{\rm {b}}$ Assuming Bond albedo $A=0.3$ and emissivity $\epsilon=1$ \citep{allesfitter-paper}. $^{\rm {c}}$ Obtained from the astrometric constraints (Sect.~\ref{SSect:Results.Astrometry}). $^{\rm {d}}$ Calculated from $\rho_\mathrm{\star,tr}$ and the stellar mass $M_{\star}$ as a consistency check, not used in calculations.}
\end{table*}

\end{document}